\documentclass[a4paper,11pt]{article}
\pdfoutput=1 
\usepackage{jcappub} 
\usepackage{amsmath,amsfonts,amssymb,mathtools}
\usepackage[numbers]{natbib}
\usepackage[T1]{fontenc} 
\usepackage{booktabs}
\usepackage{multirow}
\usepackage{xcolor}

\newcommand{\fnl}{f_{\rm NL}}
\newcommand{\be}{\begin{equation}}
\newcommand{\ee}{\end{equation}}
\newcommand{\bea}{\begin{eqnarray}}
\newcommand{\eea}{\end{eqnarray}}
\newcommand{\bdm}{\begin{displaymath}}
\newcommand{\edm}{\end{displaymath}}

\newcommand{\avg}[1]{\ensuremath{\left\langle \,#1\, \right\rangle}}
\allowdisplaybreaks

\title{\boldmath Optimal constraints on Primordial non-Gaussianity with the eBOSS DR16
quasars in Fourier space}
\author[a,1]{Marina S. Cagliari,\note{Corresponding author}}
\author[a,b]{Emanuele Castorina,}
\author[c,d,e]{Marco Bonici,}
\author[a]{Davide Bianchi}

\affiliation[a]{Dipartimento di Fisica ‘Aldo Pontremoli’, Università degli Studi di Milano, \\ Via Celoria 16, 20133 Milan, Italy}
\affiliation[b]{INFN, Sezione di Milano,
Via Celoria 16, 20133 Milan, Italy}
\affiliation[c]{Waterloo Centre for Astrophysics, University of Waterloo, Waterloo, ON N2L 3G1, Canada}
\affiliation[d]{Department of Physics and Astronomy, University of Waterloo, 200 University Ave W, Waterloo, ON N2L 3G1, Canada}
\affiliation[e]{INAF-IASF Milano, Via Alfonso Corti 12, I-20133 Milano, Italy}
\emailAdd{marina.cagliari@unimi.it}
\emailAdd{emanuele.castorina@unimi.it}
\emailAdd{mbonici@uwaterloo.ca}
\emailAdd{davide.bianchi1@unimi.it}

\begin{document}

\abstract{
We present constraints on the amplitude of local Primordial Non-Gaussianities (PNG), $f_{\rm NL}$, using  the quasar sample in the Sloan Digital Sky Survey IV extended Baryon Oscillation Spectroscopic Survey Data Release 16. We analyze the power spectrum monopole, testing for the presence of scale dependent galaxy bias induced by local PNG. Our analysis makes use of optimal redshift weights that maximize the response of the quasar sample to the possible presence of non zero PNG. We find $-4< f_{\rm NL} <27$ at 68\% confidence level, which is among the strongest bounds with Large Scale Structure data. The optimal analysis reduces the error bar by $\sim10\%$ compared to the standard one, an amount which is smaller than what was expected from a Fisher matrix analysis. This, and the reduced improvement over previous releases of the same catalog, suggest the presence of still unknown systematic effects in the data.  
If the quasars have a lower response to local PNG, our optimal constraint becomes $-23< f_{\rm NL}<21$ at 68\%, with an improvement of 30\% over standard
analyses. We also 
show how to use the optimal weights to put data-driven priors on the sample's response to 
local PNG.}


\maketitle
\flushbottom

\section{Introduction and main results} 
\label{sec:introduction}

The late time distribution of the Large Scale Structure (LSS) of the Universe is the result of the evolution, under gravitational interaction, of the set of primordial curvature perturbations. By measuring the $n$-point functions of a galaxy sample we have therefore the unique opportunity to test the statistical properties of the initial conditions of the Universe. Of particular relevance for LSS probes is the presence of possible Primordial Non-Gaussianities (PNG). The leading hypothesis for the dynamical generation of the primordial density fluctuations, Inflation (see \cite{Baumann:2009ds} for a review), offers theoretical guidance to the most generic ways PNG could arise in cosmological correlators, also indicating that PNG are generically smaller than the dominant Gaussian term.

In this work we focus on the so-called local PNG, for which the primordial gravitational potential $\Phi_P (\bf{x})$ is a non-linear function of a Gaussian field $\varphi$, $\Phi_P = \varphi + f_{\rm NL} (\varphi^2-\avg{\varphi^2})$. The amplitude of local PNG is parameterized by a single number $f_{\rm NL}$, and we immediately see that, if the primordial fluctuations are of $\mathcal{O}(10^{-5})$, local PNG are $\mathcal{O}(10^{5})$ smaller than the Gaussian term for $f_{\rm NL} =1$. Local PNG are among the most studied in the literature because they are exactly zero if the inflationary dynamics is driven by a single degree of freedom, the so-called single-field models \cite{Maldacena:2002vr,Creminelli:2004yq,Cabass:2016cgp}. A robust detection of $f_{\rm NL}$ will therefore exclude all such models and point to a more complicated inflationary sector. Conversely, multi-field models of inflation generically predict $f_{\rm NL} \sim \mathcal{O}(1)$ \cite{Senatore:2010wk,Alvarez:2014}, and could be severely constrained by a strong experimental bound. Measurements of the anisotropies of the Cosmic Microwave Background (CMB) from the Planck satellite put the stringent limit $\fnl = 0.8 \pm 5$ \cite{2020A&A...641A...9P}, and upcoming instruments are expected to reduce this error bar by another 50\% \cite{CMB-S4:2016ple}. Differently than the CMB, which is sensitive to $\fnl$ starting with the three-point function, LSS can probe PNG at the two-point, or power spectrum in Fourier space, level. As first pointed out in ref.~\cite{Dalal:2008}, the quadratic term in the definition of the primordial potential $\Phi_P$ induces a correlation between the long-wavelength gravitational field and the small scales fluctuations. The latter could very well be in the range corresponding to the formation of halos and galaxies, whose number density is therefore modulated by the large scale value of $\Phi_P$. Mathematically, we say that the large scale bias of galaxies is modified in presence of local PNG, and it reads
\begin{align}
\label{eq:bias}
    \delta_g (\mathbf{x},z) = b(z) \, \delta_m (\mathbf{x},z) + \fnl \,  b_\phi(z) \, \Phi_P (\mathbf{x}) \,
\end{align}
where $\delta_g$ is the galaxy density perturbation and $b$ is the Gaussian linear bias. The new bias coefficient $b_\phi$ parameterizes the actual response of small scale fluctuations to the presence of local PNG, and it is subject to large theoretical uncertainties due to our incomplete knowledge of galaxy formation physics \cite{Barreira:2021ueb,Barreira:2022sey}.\footnote{This point could however be turned the other way around and suggests that by carefully selecting the galaxy sample one could maximize the response, i.e. the value of $b_\phi$, to improve the constraint \cite{Castorina:2018zfk, Sullivan:2023qjr,Barreira:2023rxn}.} Via the Einstein's Equations, the presence of $\Phi_P$ in the above expression implies that, on large scales, the power spectrum acquires a distinct $k^{-2}$ feature, which is then interpreted as the smoking gun of local PNG. In this respect, knowing the value of $b_\phi$ is not a fundamental limitation, since what ultimately matters to exclude single field models is a detection rather than the actual value of $\fnl$. 
The possibility to measure local PNG with the galaxy power spectrum has spurred a tremendous amount of research activities, and all major spectroscopic and photometric instruments like the Dark Energy Spectroscopic Instrument \citep[DESI;][]{desi}, Euclid \cite{euclid}, SPHEREx \cite{spherex} and the Vera Rubin Observatory \citep{lsst}, have the search for PNG as one of their primary science goals. It also serves as an important science case for future facilities \citep{Achucarro:2022qrl}. 
Current LSS bounds are still far from the CMB one, $|\fnl| \sim \mathcal{O}(20-30)$ \cite{Castorina2019, Mueller:2022dgf,DAmico:2022gki,Cabass:2022ymb}, but are expected to improve down to $\sigma_{\fnl} \sim 1$ with current and future observations \cite{Sailer:2021yzm,Cabass:2022epm,CosmicVisions21cm:2018rfq,Braganca:2023pcp,Karagiannis:2018jdt}.\footnote{Recent work, ref.~\cite{Rezaie:2023lvi}, presents evidence of non zero $\fnl$ at more than 99\% confidence level (c.l.) with DESI imaging data. However, CMB and LSS measure local PNG on the same range of scales, which suggests a non-cosmological origin for the signal reported in ref.~\cite{Rezaie:2023lvi}.}

The main goal of this work is to provide the most stringent and robust constraints on local PNG with current data. We will use  the extended Baryon Oscillation Spectroscopic Survey (eBOSS) Data Release 16 quasar (QSO) sample \citep[DR16Q;][]{2020MNRAS.498.2354R,2020ApJS..250....8L}. 
Our analysis takes advantage of optimal signal weighting that maximizes the response of a given galaxy sample to the presence of local PNG. These weights were first derived in ref.~\cite{Castorina2019}, and are based on optimal quadratic estimators \citep{1997PhRvD..55.5895T,2000ApJ...533...19B,1998ApJ...499..555T}. The main reason to use optimal weights lies in eq.~\eqref{eq:bias}: the non-Gaussian contribution is proportional to the primordial potential, and therefore it does not evolve over time, while the linear bias term is proportional to the matter density, which grows over time. 
This suggests that, in a given sample, high-redshift objects should be given more weight than low-redshift ones, since the Gaussian piece is smaller at earlier times. As we will see in section~\ref{sec:data} in more detail, the optimal analysis downweights the Gaussian signal by $w_0 \sim b(z)D(z)$, where $D(z)$ is the linear growth factor that decreases with increasing redshift, and upweights the PNG term by $\tilde{w}\sim b_\phi$. 

The optimal redshift weighting therefore requires some prior knowledge of the response $b_\phi$ of a given sample to the presence of local PNG.  For mass selected halos, analytical models \cite{Slosar:2008hx,Biagetti:2019bnp} and simulations \cite{Biagetti:2016ywx,Barreira:2020kvh} suggest that $b_\phi \propto (b-p)$, where $p$ is a number of $\mathcal{O}(1)$, is a very good approximation to the true response. As stated above, the picture is however much more complicated for observed galaxies. In this work we will present constraints on $\fnl$ for the two values of $p$ mostly used in the literature, $p=1.0$ and $p=1.6$. The fact that the bound on local PNG depends on $p$ could raise some concern about the robustness of our results. However, as we already pointed out, what matters is only a possible detection of $\fnl$, not what the actual value is. Moreover, galaxies selection based on luminosity, color, or magnitude, correlates reasonably well with host halo mass \cite{Scoccimarro:2000gm,Yuan:2022rsc}, therefore we do not expect large deviation from the values used in this work.
Nevertheless, we will also show, in a novel application of our framework, how optimal signal weights can be used to put a data-driven prior on the value of $b_\phi$ or $p$. As mentioned above, the optimal weights $\tilde{w}$ are proportional to the response of the galaxy number density to the presence of local PNG. This implies that if the input value of $b_\phi$ we use for the weighting is very different from the true response, the optimal analysis will not improve the bound on $\fnl$ over the un-weighted case, or  will even worsen the constraints. 
As a first application of this idea, we will show in section~\ref{sec:results} that a large value of $p \gtrsim 3$ for the response of the QSOs in DR16 is not favored by the data, without relying on any numerical simulations. We expect that 
as the uncertainty on $\fnl$ reduces in the near future, our method will provide invaluable information on the most likely value of $p$ to use in the data analysis.
It should, however, be kept in mind that assuming a value of $p$ implies that different data sets cannot be combined together or with the CMB. For this reason we will also show, in appendix~\ref{sec:App}, constraints for $b_\phi \fnl$ for all the data sets used in this work. 

Our strongest bounds read
\begin{equation}
\begin{cases}
    -4 < \fnl < 27 \, ,  \quad  \;\; \,68\%\,\text{c.l.} \, ,\\
    -18 < \fnl < 42 \, ,  \quad  \; 95\%\,\text{c.l.} \, ,
\end{cases}  \text{for } p=1.0\,,
\label{eq:best-bounds_p1p0}
\end{equation}
and%
\begin{equation}
\begin{cases}
    -23 < \fnl < 21 \, ,  \quad 68\%\,\text{c.l.}  \, , \\
    -43 < \fnl < 44 \, ,  \quad 95\%\,\text{c.l.} 
\end{cases}\text{for } p=1.6\,,
\label{eq:best-bounds95}
\end{equation}
which should be compared with a standard Feldman-Kaiser-Peacock \citep[FKP;][]{1994ApJ...426...23F} analysis, see figure~\ref{fig:fnl-posterior} and table~\ref{tab:fnl} for the full results. The optimal analysis improves by $10\%$ and $30\%$ over the FKP one for the $p=1.0$ and $p=1.6$ cases respectively. It is worth stressing that the power spectrum of DR16Q catalog is, on all scales, dominated by the shot-noise, and we therefore did not expect much larger gains \cite{Castorina2019}.
Our optimal constraints are robust to the treatment of systematic effects. The bounds using a linear method to remove known foregrounds are statistically indistinguishable from the ones obtained with a non-linear algorithm based on Neural-Network \citep[NN;][]{2021MNRAS.506.3439R}. However, compared to the previous eBOSS data release \cite{Castorina2019}, the improvement in the constraint  is smaller than what was expected from the increase in volume, and it is most likely due to the presence of residual foregrounds in the maps. This could also be the reason of the more limited improvement of the optimal analysis with respect to the standard one in comparison to the improvement found in DR14 \cite{Castorina2019}.

Our results are roughly comparable with the ones in ref.~\cite{Mueller:2022dgf}, which also used the DR16Q data set. However there are a number of important differences with our analysis. First, the weights employed in ref.~\cite{Mueller:2022dgf} are defined for pair of galaxies, and cannot be automatically applied to individual galaxies, as relevant for a power spectrum analysis. To avoid imaginary weights for a single object, the Authors of \cite{Mueller:2022dgf} imposed, by hand, the positivity of the weights, which is not by itself an optimal procedure.
In our case, the weights can very well be negative, precisely in the region where the signal is: in this way the product of the weights times the signal contributes positively to the total signal-to-noise. More generally, cosmological information is contained in the galaxy fields and not in its non-linear transformations, like for example pairs of galaxies.
The other main difference with the work of \cite{Mueller:2022dgf} is in the modeling of the signal. As discussed in section~\ref{sec:window}, we think the bound of ref.~\cite{Mueller:2022dgf} is artificially tighter due to an incorrect choice for the effective redshift, $z_{\rm eff}$, at which the theoretical model is evaluated. For most applications the precise definition of $z_{\rm eff}$ does not matter, but it becomes important in searches for local PNG, where the signal is proportional to $b_{\phi}(z) \sim b(z)-p$. For the DR16Q analysis in ref.~\cite{Mueller:2022dgf}, a too high value of $z_{\rm eff}$ results in a higher linear bias $b(z)$, which artificially reduces the uncertainty on $f_{\rm NL}$ to keep the product $b_{\phi} \fnl \sim \text{constant}$. In section~\ref{sec:window} we will also clarify on this issue and on what the more accurate definition of $z_{\rm eff}$ is.

The rest of this paper is organized as follows: in section~\ref{sec:data} we present the DR16Q data set and the measurements of the power spectrum; in section~\ref{sec:method} we discuss the modeling of the QSO power spectrum, its convolution with the window function and the definition of the effective redshift; in section~\ref{sec:results} we present and discuss the constraints on $f_{\rm NL}$; section~\ref{sec:conclusions} concludes and summarizes our results.

All the codes, scripts, measurements and MonteCarlo Markov chains used in this work are freely accessible at \url{https://github.com/mcagliari/eBOSS-DR16-QSO-OQE}.

\section{Data} \label{sec:data}

\subsection{The eBOSS QSO Sample} \label{sec:sample}
\begin{figure}
    \centering
    \includegraphics[width=.7\textwidth]{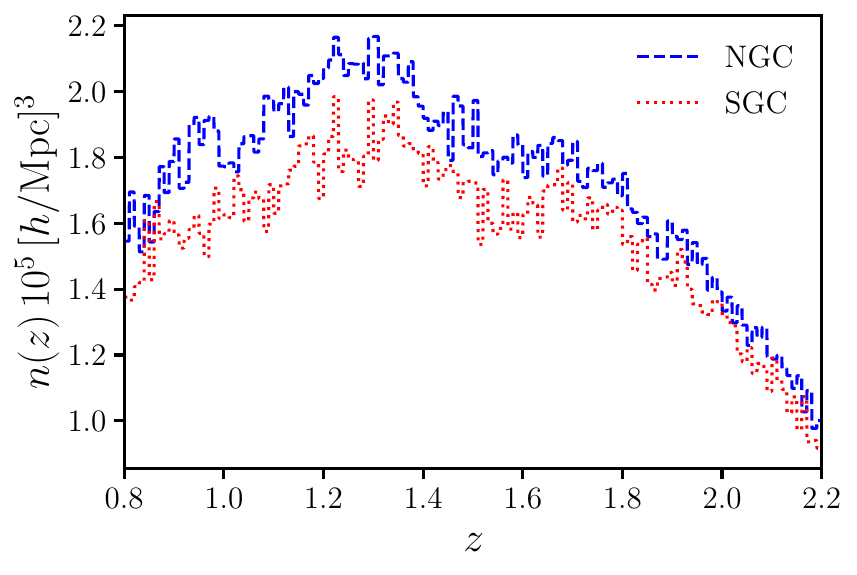}
    \caption{The quasar number density as a function of redshift for the DR16Q sample. The dashed blue line corresponds to the NGC, and the dotted red line to the SGC. The SGC has a lower density in comparison to the NGC due to the difference in mean depth in the two regions \citep{2020MNRAS.498.2354R}.}
    \label{fig:nz}
\end{figure}
In this work we use the eBOSS DR16Q sample \citep{2020MNRAS.498.2354R,2020ApJS..250....8L}. As part of the Sloan Digital Sky Survey IV (SDSS-IV) experiment \citep{2017AJ....154...28B}, the eBOSS data were acquired at the Apache Point Observatory in New Mexico.

The DR16Q sample contains $343\,708$ quasars in the redshift range $0.8 < z < 2.2$. The sample is divided into two fields of view, the North Galactic Cap (NGC), which covers an area of $2\,924 \, \text{deg}^2$, and the South Galactic Cap (SGC), with an area of $1\,884 \, \text{deg}^2$; in comparison to Data Release 14 (DR14) the area is approximately doubled. The whole sample has a volume of $\sim 20 \, (\text{Gpc}/h)^3$. The NGC has a mean density of $n \approx 1.8 \times 10^{-5} \, (\text{Mpc}/h)^{-3}$, while the SGC has a slightly lower density of  $n \approx 1.6 \times 10^{-5} \, (\text{Mpc}/h)^{-3}$.  The number densities as a function of redshift of the NGC and the SGC quasars are show in figure~\ref{fig:nz}. The number density of SGC is about $10\%$ lower than the NGC number density because of the lower mean depth of the survey in the SGC region. The data of the North and South Galactic Cap were released in two separated catalogs, each one with the corresponding random catalog. The random catalogs are $50$ times more dense than the data catalogs, and their redshift distributions are produced by sampling from the observed data redshifts \citep{2020MNRAS.498.2354R}, a procedure known as shuffling.
The use of the shuffling scheme to produce the random catalogs introduces a systematic effect called radial integral constraint \citep[RIC;][]{2019JCAP...08..036D}, the estimation and correction of which are going to be discussed and section~\ref{sec:window-ic}.

Both the data and random catalogs contain three weights for each data point. First, there are the close pairs weights, $w_{\text{cp}}$, which take into account fiber collisions. The second weights are related to the spectroscopic completeness, $w_{\text{noz}}$, and correct for the expected redshift failure rate. Third, there are the imaging systematic weights, $w_{\text{sys}}$. These weights correct for systematic effects at large angular scales, and they are therefore especially important for $f_{\text{NL}}$ measurements. In the official data release of eBOSS DR16 \cite{2020MNRAS.498.2354R}, these weights are computed with linear regression of the imaging properties and Galactic foregrounds. For the quasars, additional catalogs  were released \citep{2021MNRAS.506.3439R}, in which the imaging systematic weights were computed using neural networks. Neural networks are able to approximate non-linear functions, hence they could in principle produce a better correction than the weights computed with the linear regression. NN methods will however remove part of the signal as well, and could bias negative the constraint on local PNG. 
Hereafter we will refer to the official DR16Q catalogs as the linear weight catalog, and to the catalogs with NN systematic weights as NN weight catalog.
The completeness weight contribution to any data point is
\begin{equation}
    w_{\text{c}} = w_{\text{cp}} \, w_{\text{noz}} \, w_{\text{sys}} \, ,
\label{eq:wc}
\end{equation}
where $w_{\text{sys}}$ can either be from the linear weight catalog or the NN weight catalog. The same weighting procedure of eq.~\eqref{eq:wc} applies to the objects in the random catalogs. 

\subsection{Mocks} \label{sec:mocks}
A set of $1000$ synthetic clustering catalogs for each Galactic cap \citep{2021MNRAS.503.1149Z} was simulated using the effective Zel'dovich approximation mock method \citep[EZmock;][]{2015MNRAS.446.2621C}. The EZmock catalogs were produced assuming a flat $\Lambda$CDM cosmology with $\Omega_m = 0.307115$, $\Omega_{\Lambda} = 0.62885$, $\Omega_b = 0.048206$, $h = 0.6777$, $\sigma_8 = 0.8225$, $n_s = 0.9611$, and $f_{\text{NL}} = 0$. The mocks reproduce the two and three-point clustering statistics of DR16Q. 

In the official release of the eBOSS DR16 EZmock catalogs, three sets of mock catalogs are provided. Each set of EZmocks consists of $1000$ pairs of data and random catalogs. We refer to the first set of EZmocks as EZmock \textit{realistic}, since the data and random catalogs of this set contain all the known observational systematic effects. Each data catalog has a corresponding random catalog, whose redshift distribution is produced by shuffling the redshift of the data catalog. The $1000$ data-random pairs of the EZmock realistic set are used to estimate the covariance matrix, $\mathbf{\Sigma}$. The imaging systematic weights of the realistic EZmock are computed only with the linear regression method and not with the neural network. The covariance matrix will thus only contain information about the linear imaging weights. The other two sets are the EZmocks \textit{complete} and the EZmocks \textit{shuffled}. These two sets share the same data catalogs, which do not have any observational systematic effect, while the random catalog redshift distributions were produced in different ways. In the EZmock shuffled set each data catalog has a corresponding random catalog the redshift distribution of which is produced with the shuffling scheme. The EZmock complete set only has two random catalogs, one for each Galactic cap. The redshift distributions of these random catalogs were sampled from the same $n(z)$ interpolation used for the EZmock data catalogs. We use the EZmock complete and shuffled sets to estimate the RIC effect. 

\subsection{Power spectrum estimation} \label{sec:power_spec}
\begin{figure}
    \centering
    \includegraphics[width=.7\textwidth]{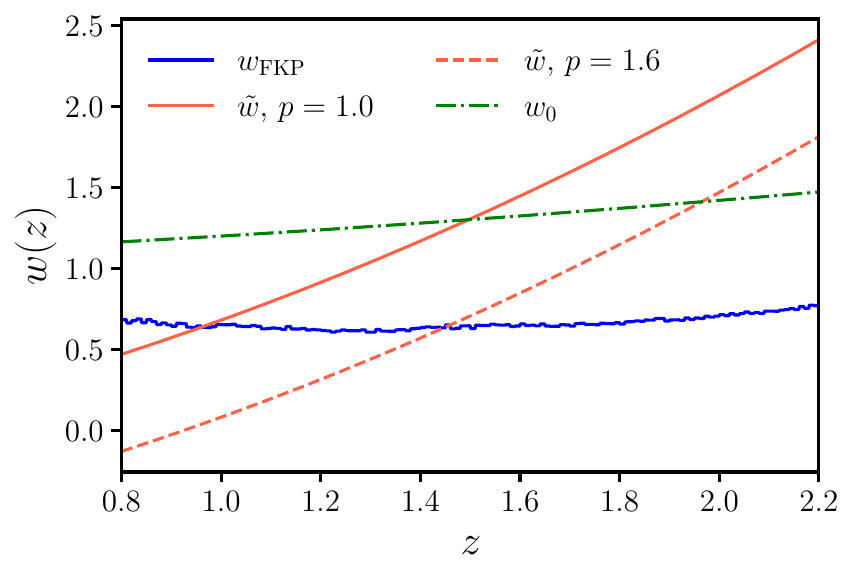}
    \caption{Weights as a function of redshift for the eBOSS DR16Q. The solid blue line corresponds to the FKP weights, eq.~\eqref{eq:wfkp}; its almost flat behavior as a function of redshift is due to $n(z) \, P_{\text{fid}} \ll 1$. The optimal weights to estimate $f_{\text{NL}}$ are shown in red and dot-dashed green. The solid and dashed red lines respectively correspond to $\tilde{w}(z)$ for $p=1.0$ and $p=1.6$. These weights have a clear dependence on the quasar response to the $f_{\text{NL}}$ signal.}
    \label{fig:wz}
\end{figure}
As shown in \cite{Castorina2019}, the optimal power spectrum is the cross-correlation of two different fields produced by weighting the underlying catalog. 
To arrive at the power spectrum estimator we start from the two quasar density fields, 
\begin{equation}
    \tilde{F}(\mathbf{r}) = \tilde{w}_{\text{tot}} \, \left[ w^{\text{qso}}_{\text{c}} \, n_{\text{qso}}(\mathbf{r}) - \alpha_{\text{s}} w^{\text{s}}_{\text{c}} \, n_{\text{s}}(\mathbf{r}) \right] \, , \quad F_0(\mathbf{r}) = w_{\text{tot}, \, 0} \, \left[ w^{\text{qso}}_{\text{c}} \, n_{\text{qso}}(\mathbf{r}) - \alpha_{\text{s}} w^{\text{s}}_{\text{c}} \, n_{\text{s}}(\mathbf{r}) \right] \, ,
    \label{eq:qso-fields}
\end{equation}
where $n_{\text{qso}}$ and $n_{\text{s}}$ respectively are the number density of the quasar sample and the corresponding random catalog, $w^{\text{qso}}_{\text{c}}$ and $w^{\text{s}}_{\text{c}}$ are the completeness weights from eq.~\eqref{eq:wc} for the quasars and the randoms. The total weights, $\tilde{w}_{\text{tot}}$ and $w_{\text{tot}, \, 0}$, are the product of the FKP weights \citep{1994ApJ...426...23F},
\begin{equation}
    w_{\text{FKP}}(z) = \frac{1}{1 + \bar{n}(z) \, P_{\text{fid}}} \, ,
    \label{eq:wfkp}
\end{equation}
and the optimal weights for a $f_{\text{NL}}$ measurements with power spectrum data \citep{Castorina2019},
\begin{equation}
    \tilde{w}(z) = b(z) - p \, , \quad w_0(z) = D(z) \, \left( b(z) + \frac{f(z)}{3} \right) \, .
    \label{eq:opt-weights}
\end{equation}
Therefore the total weights read
\begin{equation}
    \tilde{w}_{\text{tot}}(z) = w_{\text{FKP}}(z) \, \tilde{w}(z) \, , \quad w_{\text{tot}, \, 0}(z) = w_{\text{FKP}}(z) \, w_0(z) \, .
    \label{eq:wtot}
\end{equation}
In eq.~\eqref{eq:wfkp}, $\bar{n}(z)$ is the mean density as a function of redshift, and $P_{\text{fid}} = 3 \times 10^4 \, (\text{Mpc}/h)^3$, which corresponds to the expected power on the scales effected by PNG in the sample. In eq.~\eqref{eq:opt-weights},  $b(z)$ is the fiducial value of the QSO bias model~\cite{2017JCAP...07..017L}, 
\begin{equation}
    b(z) = 0.278 \, \left( (1 + z)^2 - 6.565 \right) + 2.393 \, ,
    \label{eq:bz}
\end{equation}
and $D(z)$ and $f(z)$ are respectively the growth factor and growth rate as functions of redshift. 
As mentioned in section~\ref{sec:introduction}, in this work we use $p=1.0$ and $p=1.6$. The dependence on redshift of the weights defined in eq.~\eqref{eq:wfkp} and eq.~\eqref{eq:opt-weights} is plotted in figure~\ref{fig:wz}. It shows the difference between the FKP weighting scheme, which is almost constant in redshift, as $n(z) P_{\text{fid}} \ll 1$, and the optimal weights, which have a strong dependence on redshift. Finally, the factor $\alpha_{\text{s}}$ in eq.~(\ref{eq:wfkp}) is defined as
\begin{equation}
    \alpha_{\text{s}} = \frac{\sum^{\text{qso}} w_{\text{c}}}{\sum^{\text{s}} w_{\text{c}}} \, ,
    \label{eq:alpha}
\end{equation}
and it properly normalizes the number density of the random catalog.

Following ref.~\citep{2006PASJ...58...93Y}, we write the monopole of the cross-correlation between the two weighted fields in eq.~\eqref{eq:qso-fields} as
\begin{equation}
    \widetilde{P}_0 (k) = A^{-1}_0 \, \int \frac{\text{d}\Omega_k}{4 \pi} \left[ \int \text{d}\mathbf{r}_1 \, \tilde{F}(\mathbf{r}_1) \, e^{i\mathbf{k} \cdot \mathbf{r}_1} \, \int \text{d}\mathbf{r}_2 \, F_0(\mathbf{r}_2) \, e^{-i \mathbf{k} \cdot \mathbf{r}_2} \mathcal{L}_0(\mathbf{\hat{k}} \cdot \mathbf{\hat{r}}_2) \right] - S_0 \, ,
    \label{eq:P0-yamamoto}
\end{equation}
where $\mathcal{L}_0$ is the first Legendre polynomial. The normalization factor $A_0$ and the shot noise contribution, $S_0$, are respectively defined as
\begin{equation}
    A_0 = \int \text{d}\mathbf{r} \, w_{\text{tot}, \, 0}(\mathbf{r}) \, \tilde{w}(\mathbf{r}) \, \left[ w_{\text{c}} \, n_{\text{qso}}(\mathbf{r}) \right]^2 \, ,
    \label{eq:int-A}
\end{equation}
\begin{equation}
    S_0 = A^{-1}_0 \, \int \text{d} \mathbf{r} \, w_{\text{c}} \, n_{\text{qso}}(\mathbf{r}) \, \left( w_{\text{c}}(\mathbf{r}) + \alpha_{\text{s}} \right) \, w_{\text{tot}, \, 0}(\mathbf{r}) \, \tilde{w}(\mathbf{r}) \, \mathcal{L}_0(\mathbf{\hat{k}} \cdot \mathbf{\hat{r}}) \, .
    \label{eq:int-S}
\end{equation}

We calculate the monopole of the power spectrum using \texttt{nbodykit} \citep{2018AJ....156..160H}, which implements eq.~\eqref{eq:P0-yamamoto} as follows \cite{Bianchi:2015oia,Hand:2017irw},
\begin{equation}
    \widetilde{P}_0(k) = A^{-1}_0 \int \frac{\text{d}\Omega_k}{4 \pi} \tilde{F}(\mathbf{k}) \, F_0(-\mathbf{k}) \, ,
    \label{eq:P0-nb}
\end{equation}
with
\begin{align}
    F_0(\mathbf{k}) & = \int \text{d}\mathbf{r} \, F_0(\mathbf{r}) \, e^{i \mathbf{k} \cdot \mathbf{r}} \mathcal{L}_0(\hat{\mathbf{k}} \cdot \hat{\mathbf{r}}) \nonumber \\
    & = 4 \pi \, Y_{00}(\hat{\mathbf{k}}) \int \text{d}\mathbf{r} \, F_0(\mathbf{r}) \, Y^*_{00}(\hat{\mathbf{r}}) \, e^{i \mathbf{k} \cdot \mathbf{r}} \, ,
    \label{eq:F0}
\end{align}
where $Y_{00}$ is the first spherical harmonic. The normalization and the shot noise are computed as discrete sums over the quasars and the randoms. The normalization is
\begin{equation}
    A_0 = \alpha_{\text{s}} \sum^{N_{\text{s}}}_i n_{\text{s}}(\mathbf{r}_i) \, w_{\text{c}}(\mathbf{r}_i) \, w_{\text{tot}, 0}(\mathbf{r}_i) \, \tilde{w}_{\text{tot}}(\mathbf{r}_i) \, .
    \label{eq:discr-As}
\end{equation}
The shot noise contribution becomes
\begin{equation}
    S_0 = A^{-1}_0 \left[ \sum^{N_{\text{qso}}}_i w^2_{\text{c}}(\mathbf{r}_i) \, w_{\text{tot}, 0}(\mathbf{r}_i) \, \tilde{w}_{\text{tot}}(\mathbf{r}_i) + \alpha^2_{\text{s}} \sum^{N_{\text{s}}}_i w^2_{\text{c}}(\mathbf{r}_i) \, w_{\text{tot}, 0}(\mathbf{r}_i) \, \tilde{w}_{\text{tot}}(\mathbf{r}_i)\right] \, .
    \label{eq:discr-S}
\end{equation}
To compute the power spectrum estimator in eq.~\eqref{eq:P0-nb} we use a mesh of $512^3$ cells. The quasars and the random objects are projected onto the mesh using a triangular shaped cloud interpolation \citep{1981csup.book.....H}. In this interpolation each quasar and random is weighted by both its completeness and total weight and we assume Planck \citep{Planck18} as fiducial cosmology. The power spectrum is estimated on a logarithmic grid from $k_{\text{min}} = 3.75 \times 10^{-3} \, (\text{Mpc}/h)^{-1}$ to $k_{\text{max}} = 2.23 \times 10^{-1} \, (\text{Mpc}/h)^{-1}$ for NGC, and $k_{\text{max}} = 2.78 \times 10^{-1} \, (\text{Mpc}/h)^{-1}$ for SGC.

A final remark about the normalization of the power spectrum estimator in eq.~\eqref{eq:P0-nb} and the shot noise contribution in eq.~\eqref{eq:discr-S}. Since eq.~\eqref{eq:discr-As} is just an approximation of the exact definition in eq.~\eqref{eq:int-A}, it has been pointed out that this could lead to biased constraints on cosmological parameters \cite{deMattia:2020fkb}. We therefore decided to re-normalize the measured power spectra by the limit, at small separation, of the monopole of the window function, $Q_0(0)$, see section~\ref{sec:window}. The final estimator of the power spectrum is
\begin{equation}
    \hat{P}_0(k) = \frac{A_0}{Q_0(0)} \left( \widetilde{P}_0(k) - S_0 \right) \, .
    \label{eq:P0-norm}
\end{equation}
We note that we used, both for the linear and NN catalog, the value of $Q_0(0)$  computed from the randoms of the linear catalog, since the two catalogs are expected to have the same response at small scales.

\begin{figure}
    \centering
    \includegraphics[width=1\textwidth]{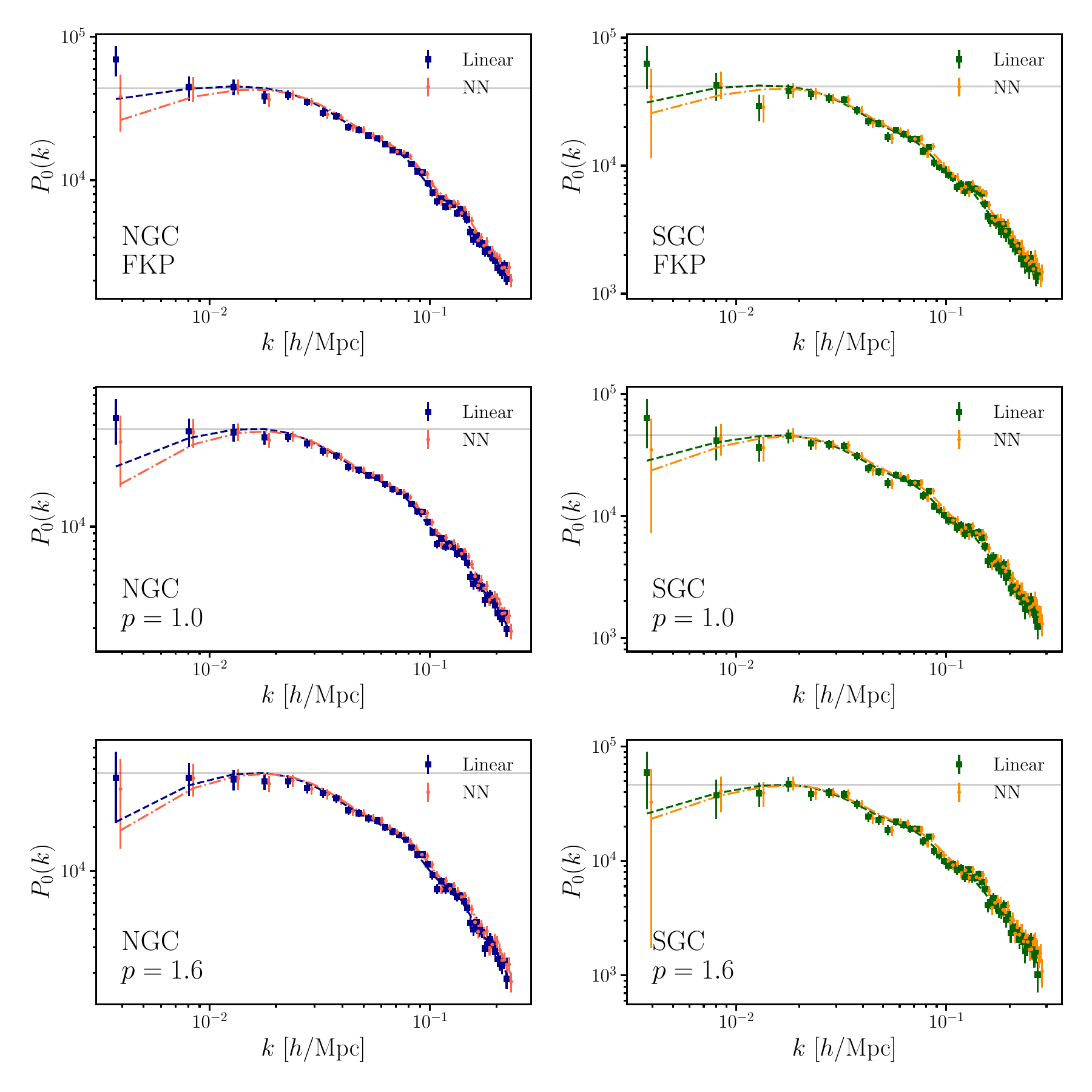}
    \caption{Observed power spectra and best fit model for the NGC (right column), and SGC (left column), and for the three weighting scheme: FKP (top row), optimal weights with $p=1.0$ (middle row), and with $p=1.6$ (bottom row). For the NGC (SGC) blue (green) dots and the dashed line correspond to the linear catalog results, red (yellow) squares and dash-dotted line to the NN catalog. The horizontal gray line marks the turn-around point in the power spectrum. To better visualize the power spectra error bars, the NN $P(k)$ was shifted of $5\%$.}
    \label{fig:Pk-obs}
\end{figure}
In figure~\ref{fig:Pk-obs} we show the NGC and SGC (left and right columns) observed power spectra of the linear and NN catalog (dots and squares). The three rows correspond to the different weights used to estimate the $\hat{P}_0(k)$, from top to bottom the columns corresponds to the FKP weights, the optimal weight with $p=1.0$, and with $p=1.6$. We also plotted the best fit model for the linear (dashed line) and NN (dash-dotted line) catalogs (see section~\ref{sec:results}). The best fit of the FKP weight case is for the model with $p=1.6$. For the NN catalog power spectra, the band powers have been shifted by $5\%$ along the $k$-axis for a better visualization of the corresponding error bars. The observed power spectra estimated from the two catalogs differ only in the first two bins. The NN catalog power spectra have less power in the first bin than their linear counterparts. The excess of power in the linear catalog power spectra is expected to be related to large scale systematic effect that the linear regression weights are not able to correct. The solid horizontal line in each panel shows the amplitude of the power spectrum at its peak, and it serves to guide the eye to the fact that the optimal weighted measurements are larger than the corresponding FKP ones. This was expected because the optimal weights upweight high redshift galaxies (see figure~\ref{fig:wz}), which have a higher bias and therefore a larger clustering amplitude.

\subsection{Window functions} \label{sec:window}
The window function, $W(\mathbf{s})$, represents the footprint on the sky and the redshift selection function of the survey. It is an essential ingredient to compare a power spectrum model with the observed power spectrum.

In order to evaluate the model of the observed power spectrum we will need the multipoles of the window function (see section~\ref{sec:window-ic}), defined as follows,
\begin{align}
    Q_{\ell}(s) & \equiv (2 \ell + 1) \, \int \text{d} \Omega_s \int \text{d}^3 \mathbf{s}_1 \, W(\mathbf{s}_1) \, W(\mathbf{s} + \mathbf{s}_1) \, \mathcal{L}_{\ell}(\hat{\mathbf{s}}_1 \cdot \hat{\mathbf{s}}) 
     \equiv \int \text{d} s_1 \, s_1^2 \, Q_{\ell}(s; s_1) \, .
    \label{eq:Ql-def}
\end{align}
To compute the window function multipoles we use the pair counting approach introduced in ref. \citep{2017MNRAS.464.3121W}. First, with \texttt{nbodykit} we calculate the weighted pair counts of the random catalogs as function of the three dimensional separation and the cosine of the line-of-sight angle, $RR^{\text{w}}(s, \mu)$. That is done by cross-correlating the random catalog weighted by $w_{\text{c}} \, \tilde{w}_{\text{tot}}$, and the random catalog weighted by $w_{\text{c}} \, w_{\text{tot}, \, 0}$. Second, we compute its multipoles,
\begin{equation}
    RR^{\text{w}}_{\ell}(s) = (2 \ell + 1) \int \text{d}\mu \, RR^{\text{w}}(s, \mu) \, \mathcal{L}_{\ell}(\mu) \, .
    \label{eq:RR-multipoles}
\end{equation}
Finally, in order to obtain the window function multipoles the quantity above needs to be normalized to take into account the width of the shell over which the pair counting is performed and the density of the random catalog in comparison to the data catalog. The window function multipole $\ell$ is finally defined as
\begin{equation}
    Q_{\ell}(s) = \frac{RR^{\text{w}}_{\ell}(s)}{4 \pi \, s^3 \, \text{d}\ln{s}} \frac{\left( \sum^{\text{qso}} w_{\text{c}} \right)^2 - \sum^{\text{qso}} w^2_{\text{c}}}{\left( \sum^{\text{s}} w_{\text{c}} \right)^2 - \sum^{\text{s}} w^2_{\text{c}}} \, ,
    \label{eq:window-l}
\end{equation}
with $\text{d}\ln{s} = \frac{s_{n+1} - s_n}{s}$, where $s$ is the center of the $n$-th separation bin. We stress that each set of optimal weights requires its own multipoles of the window function $Q_\ell(s)$.

To reduce the computational time of the pair counting algorithm we divided the random catalogs into five subsets and computed $RR^{\text{w}}(s, \mu)$ for each of them. These subsets are $10$ times denser than the data catalog. We calculated the window function multipoles of each random subset using eq.~\eqref{eq:window-l}, with the caveat that the sum over the random $\sum^{\text{s}}$ is now over the subset. The final $Q_\ell(s)$ is the mean of the five subsets.

\section{Analysis Methods}
\label{sec:method}
\subsection{The Power spectrum Model}
\label{sec:model}
To model the quasar power spectrum in redshift-space we use linear theory. Linear theory is enough to make the prediction for two reasons: first, the local $f_{\text{NL}}$ signal is at low $k$, where structure is still growing with a linear regime. Second, the smaller scales of this sample are dominated by redshift error, which dominates over the non-linearities. We write the power spectrum model as follows,
\begin{equation}
    P_{\text{qso}}(k, \mu;z) = G(k, \mu; \sigma_{\text{FoG}})^2 \, \left[ b_{\text{tot}}(k;z) + f(z) \, \mu^2 \right]^2 \, P_{m}(k;z) + N \, ,
    \label{eq:P-qso}
\end{equation}
where $P_{m}$ is the matter power spectrum in real-space, $N$ is the residual shot noise free parameter, and $f(z)$ is the growth rate. The total quasar bias includes the PNG,
\begin{equation}
    b_{\text{tot}}(k;z) = b_1 + \Delta b = b_1 + f_{\text{NL}} \, (b_1 - p) \, \tilde{\alpha}(k;z) \, ,
    \label{eq:b-tot}
\end{equation}
where $b_1$ is the quasar linear bias, and $\tilde{\alpha}(k;z)$ is
\begin{equation}
    \tilde{\alpha}(k;z) = \frac{3 \, \Omega_m \, H^2_0 \, \delta_{\text{c}}}{c^2 \, k^2 \, T(k) \, D(z)} \, .
    \label{eq:alphatilde}
\end{equation}
In eq.~\eqref{eq:alphatilde}, $\delta_{\text{c}} = 1.686$ is the critical density in the spherical collapse in a Einstein-De Sitter Universe, $\Omega_m$ is the matter density parameter, and $H_0$  the Hubble parameter, both at $z=0$, and $c$ is the speed of light. Then, $T(k)$ is the matter transfer function normalized to 1 at low-$k$, and $D(z)$ the growth factor normalized to $(1 + z)^{-1}$ in the matter dominated era. Finally, the damping of the power spectrum due to nonlinear redshift-space distortions is included with a Lorentzian function,
\begin{equation}
    G(k, \mu ; \sigma_{\text{FoG}}) = \left[ 1 + \frac{(k \, \mu \, \sigma_{\text{FoG}})^2}{2} \right]^{-1} \, ,
    \label{eq:G}
\end{equation}
where $\sigma_{\text{FoG}}$ accounts for both the typical velocity dispersion of QSOs, as well as their redshift error, which  is estimated to be $\sigma_z = 300 \, \text{km s}^{-1}$ for DR16Q, with no significant dependence on the redshift \citep{2020ApJS..250....8L}.

The quasar power spectrum multipoles are then easily computed 
\begin{equation}
    P_{\ell, \, \text{qso}}(k; z) = \frac{2 \ell + 1}{2} \int_{-1}^1 \text{d}\mu \, P_{\text{qso}}(k, \mu; z) \, \mathcal{L}_{\ell}(\mu) \, .
    \label{eq:Pl-qso}
\end{equation}
To evaluate the cosmological quantities in eq.~\eqref{eq:P-qso} and eq.~\eqref{eq:alphatilde} we assume a Planck fiducial cosmology  \citep{Planck18}, and fix the redshift to an effective value. In the following section we will discuss how the effective redshift is defined and computed. We calculate the cosmological functions with \texttt{classy}, the Python wrapper of the \texttt{CLASS} CMB Boltzmann solver \citep{2011JCAP...07..034B}.

\subsection{Convolution with the window function and the effective redshift} \label{sec:window-ic}

The ensemble average of the power spectrum estimator in eq.~\eqref{eq:P0-yamamoto} is
\begin{equation}
    \langle \hat{P}_0(k)\rangle =  \sum_{\ell, \, L} \begin{pmatrix}
\ell & L & 0\\
0 & 0 & 0
\end{pmatrix}^2 \int \text{d} s \, s^2  j_0(ks) \int \text{d}s_1 \, s_1^2 \, \xi_{\ell}(s;s_1(z)) \, Q_L(s;s_1(z)) \, ,
    \label{eq:PA}
\end{equation}
where $Q_L(s; s_1)$ is defined in eq.~\eqref{eq:Ql-def}, $\xi_{\ell}(s;s_1)$ is the multipole $\ell$ of the QSO correlation function, $j_A(ks)$ is the spherical Bessel function of order $A$, and $\big(\begin{smallmatrix}
  \ell & L & 0\\
0 & 0 & 0
\end{smallmatrix}\big)$ is a Wigner 3-$j$ symbol.

\begin{figure}
    \centering
    \includegraphics[width=.8\textwidth]{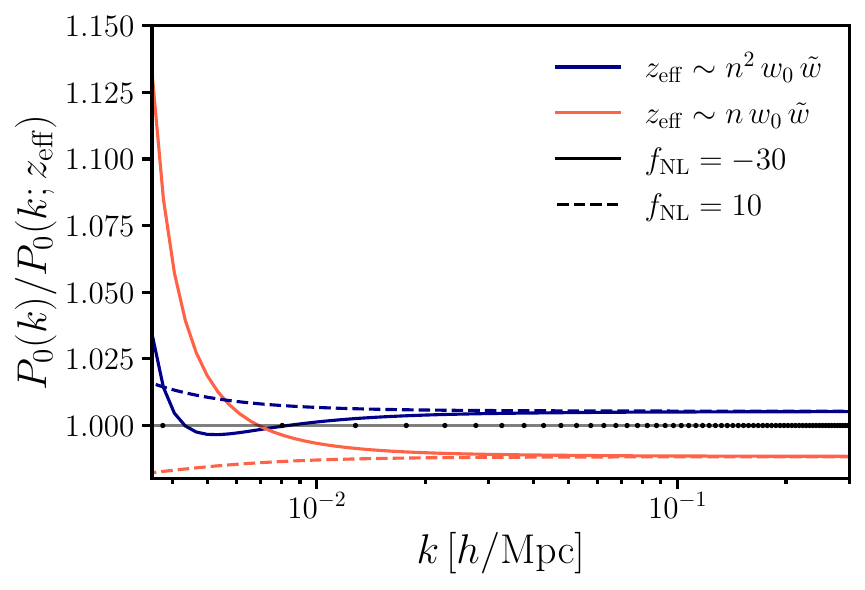}
    \caption{The accuracy of the effective redshift approximation. In red, and for two different values of $f_{\rm NL}$, the plot shows the ratio between the theoretical model integrated over the full radial selection function of the DR16Q sample and the same model evaluated at a $z_{\rm eff}$ defined in eq.~(\ref{eq:zeff}), see table~\ref{tab:zeff}. The accuracy of the other most common definition of $z_{\rm eff}$ in the literature is shown with blue lines.  The black points corresponds to the effective wavenumbers of the measurements of the QSO power spectrum.}
    \label{fig:zeff-approx}
\end{figure}

In eq.~\eqref{eq:PA} the redshift evolution of the signal is taken into account by $s_1(z)$ in $\xi_{\ell}(s;s_1)$, which should be integrated against the $Q_L(s;s_1)$ for a proper model comparison. This could be a time consuming step if repeated for every point in the parameter space exploration, and it is therefore often approximated. Noticing  that in most applications the correlation function is factorizable in time and space, $\xi(s,z) \sim g(s) \, h(z)$,  a possible way to speed up the computation of the theoretical model is to separate the integrals over $s$ and $s_1$. An even more useful approximation is to assume that the model can be evaluated only at some effective redshift, $z_{\rm eff}$, defined by the radial selection function. The expression for the power spectrum then simplifies to 

\begin{equation}
    P_0(k; z_{\text{eff}}) = \sum_{\ell, \, L} \begin{pmatrix}
\ell & L & 0\\
0 & 0 & 0
\end{pmatrix}^2 \int \text{d} s \, s^2  \, j_0(ks) \, \xi_{\ell}(s;z_{\text{eff}}) \, Q_L(s) \, ,
    \label{eq:PA-zeff}
\end{equation}
with the multipole of the correlation function of the random catalog $Q_L(s)$ defined in eq.~(\ref{eq:Ql-def}). The final integral is a simple Hankel Transform that can be computed quite efficiently \cite{2017MNRAS.464.3121W}.

\begin{table}%[ht!]
    \centering
    \begin{tabular}{cccc} \toprule 
         &  FKP & $p=1.0$ & $p=1.6$ \\ \midrule
         NGC & 1.49 & 1.65 & 1.76 \\
         SGC & 1.50 & 1.66 & 1.76 \\ \bottomrule
    \end{tabular}
    \caption{The effective redshift for the different weights and the two sky region. The optimal weights increase the $z_{\text{eff}}$ of the sample for $p=1.0$ and $p=1.6$.}
    \label{tab:zeff}
\end{table}

The question becomes, then, what the most accurate definition of $z_{\rm eff}$ is. The estimator of the power spectrum in eq.~(\ref{eq:P0-yamamoto}), and therefore the multiples of the window functions as well, contains two powers of the radial selection function, which suggests the following definition of $z_{\rm eff}$ for a sample with given $n(z)$ and weights $w(z)$,
\begin{align}
 z_{\text{eff}} = \frac{\int \text{d} z \, n(z)^2 \,  [\chi(z)^2/H(z)] \, w(z)^2 \, z}{\int \text{d} z \, n(z)^2 \, [\chi(z)^2/H(z)]\,  w(z)^2} \, ,
    \label{eq:zeff-generic}
\end{align}
where $\chi(z)$ is the comoving distance and $H(z)$ is the Hubble parameter. In practice, the integral above can be estimated via Monte-Carlo methods as
\begin{equation}
    z_{\text{eff}} = \frac{\sum^{\text{qso}} z \, n(z) \, w_{\text{c}}^2 \, w_{\text{FKP}}(z)^2 \, \tilde{w}(z) \, w_0(z)}{\sum^{\text{qso}} n(z) \, w_{\text{c}}^2 \, w_{\text{FKP}}(z)^2 \, \tilde{w}(z) \, w_0(z)}\, ,
    \label{eq:zeff}
\end{equation}
or with the analogous expression written in terms of the random catalog. The values of $z_{\rm eff}$ for the samples used in this work are shown in table~\ref{tab:zeff}. We see that the optimal weighting increases $z_{\rm eff}$, since it gives more weight to high redshift objects, which have a higher response to the presence of PNG.

The accuracy of our definition of $z_{\rm eff}$ is presented in figure~\ref{fig:zeff-approx}, with the blue lines showing the ratio between the power spectrum model fully integrated over redshift and the monopole evaluated at $z_{\rm eff}$. The dashed line corresponds to $f_{\rm NL} = 10$, while the continuous one to $f_{\rm NL} = - 30$, with $p=1.6$ in both cases. The black points on the horizontal axis show the effective values of the wavenumbers of the measurements. We find that our approximation is sub-percent accurate at high-$k$, and better than 2.5\% accurate on very large scales, thus much smaller than the sample variance of the measurements. 

On the other hand, the DR16Q analysis of ref.~\cite{Mueller:2022dgf} adopts a definition of $z_{\rm eff}$ with one less power of $n(z)$ than eq.~(\ref{eq:zeff}).\footnote{The published version of ref. \cite{Mueller:2022dgf} contains the following definition of the effective redshift, right below their eq.~(11), $z_{\rm eff} = \sum_i z_i \, w_{\rm tot}/\sum_i w_{\rm tot}$, where $w_{\rm tot}^2 = w_{\rm FKP}^2 \, w_{\rm c}^2 \, |\tilde{w} \, w_0|$. This is a typo, as the analysis of ref.~\cite{Mueller:2022dgf} actually used $z_{\rm eff} = \sum_i z_i \, w_{\rm tot}^2/\sum_i w_{\rm tot}^2$, corresponding to the red set of curves in fig.~\ref{fig:zeff-approx}. We thank Eva-Maria Mueller for correspondence about this point.} This choice produces the red set of curves in figure~\ref{fig:zeff-approx}, which we find are more than a percent off at high-$k$, a number that could become significant over many data points, and more than 10\% inaccurate at large scales.  
In particular, ref.~\cite{Mueller:2022dgf} reports $z_{\rm eff} = 1.83$ for the weights optimized with $p=1.6$, a number significantly higher than our $z_{\rm eff} = 1.76$. At fixed value of $f_{\rm NL}$, the product $b_\phi \, f_{\rm NL}  $ is 10\% larger at $z = 1.83$ than at $z=1.76$. This suggests that the authors of \cite{Mueller:2022dgf} would have at least gotten a 10\% weaker constraint on $f_{\rm NL}$, had they used the more accurate definition of $z_{\rm eff}$ in eq.~(\ref{eq:zeff-generic}).

Finally, we write the convolution of the window function in the more convenient form
\begin{align}
    P_0(k; z_{\text{eff}}) & = \sum_{\ell, \, L} i^{\ell} \begin{pmatrix}
\ell & L & 0\\
0 & 0 & 0
\end{pmatrix}^2 \int \frac{\text{d} q}{2 \pi^2} \, q^2 \, P_{\ell, \, \text{qso}}(q; z_{\text{eff}}) \int \text{d} s \, s^2 j_0(ks) \, j_{\ell}(qs) \, Q_L(s) \\
    & = \sum_{\ell, \, L} i^{\ell} \begin{pmatrix}
\ell & L & 0\\
0 & 0 & 0
\end{pmatrix}^2 \int \frac{\text{d} q}{2 \pi^2} \, q^2 \, P_{\ell, \, \text{qso}}(q; z_{\text{eff}}) \, Q_{\ell, \, L}(k, q) \, ,
   \label{eq:P0-qso}
\end{align}
where we defined
\begin{equation}
    Q_{\ell, \, L}(k, q) = \int \text{d} s \, s^2 j_0(ks) \, j_{\ell}(qs) \, Q_L(s) \, .
    \label{eq:QlLkp}
\end{equation}
The integral in eq.~(\ref{eq:P0-qso}) is then evaluated as a simple matrix multiplication. This  choices allows to never compute the correlation function multipoles, which are formally divergent in the presence of local PNG.

The multipoles of the window function corresponding to the optimal weights with $p=1.0$ are shown in figure~\ref{fig:Ql-model}, left panel. In our model we use only the even multipoles up to $\ell =4$, and neglect possible odd ones \cite{Beutler:2018vpe}.\footnote{Wide angle effects and other projection effects are negligible for the DR16Q volume \citep{Castorina:2017inr,Castorina2018,Castorina:2021xzs,Beutler:2018vpe}.}

\subsection{The Integral Constraint}
The final step to model the observed power spectrum is to correct the convolved power spectrum with the integral constraint effects. 
Integral constraint effects arise when the survey selection function is estimated from the data themselves \citep{2019JCAP...08..036D}. When the observed galaxy mean density is used as the true cosmological mean, fluctuations over the whole survey average to zero and this causes a suppression of power at large scales. This is the so called Global Integral Constraint \citep[GIC;][]{1991MNRAS.253..307P,2017MNRAS.464.3121W}. On the other hand, an additional radial integral constraint is produced when the radial $n(z)$ is inferred from the data \cite{2019JCAP...08..036D}. That is the case for eBOSS data, where the random catalog redshift distribution is obtained by shuffling the data redshift distribution. The RIC also causes a suppression of the large scale fluctuations along the line-of-sight.

\begin{figure}
    \centering
    \includegraphics[width=.5\textwidth]{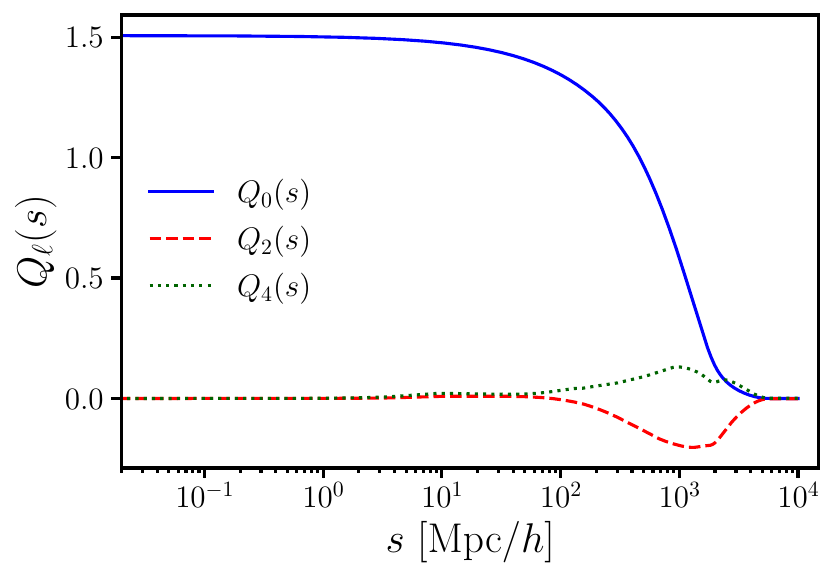}\includegraphics[width=.5\textwidth]{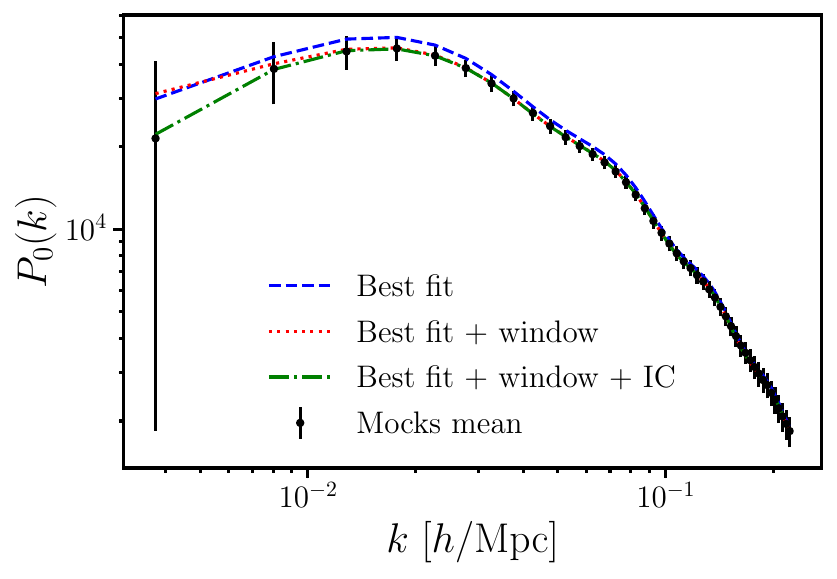} 
    \caption{\emph{Left:} multipoles of the window function in configuration space. The blue solid line, red dashed, and dotted green respectively are the monopole, quadrupole, and hexadecapole of the window function. \emph{Right}: the effect of the different components of the observed power spectrum model compared to the mean power spectrum of the EZmock realistic catalogs (black circles with error bars). The blue dashed line is the monopole of the best fit model, the dotted red line the best fit model convolved with the window function, and the dash-dotted green line is the observed power spectrum corrected with the integral constraint as in eq.~\eqref{eq:P0-IC}.}
    \label{fig:Ql-model}
\end{figure}

The global integral constraint depends on the Hankel transform, $| \tilde{W}_{\ell}(k) |^2$, of the window function multipole $Q_{\ell}(s)$. The Hankel transforms are normalized so that $| \tilde{W}_0(0) |^2 = 1$ \citep{2017MNRAS.464.3121W}. To correct for the RIC we need to estimate the effect that the shuffling of the random produces on the measured power spectrum. To do so we used both the complete and shuffled EZmocks. First, we compute the mean of the power spectra of the complete EZmocks, $\bar{P}_{\text{c}}(k)$, and the mean of the power spectra of the shuffled EZmock, $\bar{P}_{\text{r}}(k)$. Then the radial integral constraint correction is defined via
\begin{equation}
    W_{\text{RIC}}(k) = \frac{\bar{P}_{\text{c}}(k) - \bar{P}_{\text{r}}(k)}{\bar{P}_{\text{c}}(k)} \, .
    \label{eq:w-ric}
\end{equation}
Given $|W(k)|^2$ and $W_{\text{RIC}}(k)$ the final expression for the  power spectrum is
\begin{equation}
    P_0^{\text{IC}}(k) = P_0(k) - P_0(0) \, |W(k)|^2 - P_0(k) \, W_{\text{RIC}}(k) \, ,
    \label{eq:P0-IC}
\end{equation}
where we dropped the dependence on the effective redshift for brevity. In figure~\ref{fig:Ql-model} we present the effect of the different components of the observed power spectrum model and compare it with the mean power spectrum of the EZmock realistic catalogs. First, the plot shows the importance of the window convolution (dotted red line), which removes power at $k \lesssim 5 \times 10^{-2} \, (\text{Mpc}/h)^{-1}$. Second, the integral constraint correction (dash-dotted green line) only affects the first bin, but its well within the $68\%$ error bars. 

\subsection{Parameter estimation} \label{sec:parameters}

To estimate the posterior distribution of the parameters $\boldsymbol{\theta}$ of our model, $\mathbf{T}(\boldsymbol{\theta})$, given our data vector $\mathbf{D}$, we assume a multi-variate Gaussian likelihood,
\begin{equation}
    \mathcal{L}(\mathbf{D}|\boldsymbol{\theta}, \mathbf{\Sigma}) \propto \exp \left( - \frac{1}{2} \, \sum_{i j} (D_i - T_i(\boldsymbol{\theta})) \, \Sigma^{-1}_{ij} (D_j - T_j(\boldsymbol{\theta})) \right) \, .
    \label{eq:likelihod}
\end{equation}
The model parameters $\boldsymbol{\theta}$ are three for each field of view: $\sigma_{\text{FoG}}$, $b_1$ and $N$ (see section~\ref{sec:model}), and $f_{\text{NL}}$. The rest of the cosmology is fixed to the Planck best fit values \citep{Planck18}. Therefore, when fitting the data of one sky patch (single field analysis) the total number of free parameters is four, and when fitting the two fields of view data (joint analysis) the free parameters are seven. In the case of the joint analysis $f_{\text{NL}}$ is common for the two fields, while the other three parameters of the model are unique for each field of view, for a total of six parameters. We assume a uniform prior distribution for all the parameters, with the following bounds
\begin{equation}
\begin{split}
    f_{\text{NL}} & \in [-500, 500] \, , \\
    b_1 & \in [0.1, 6] \, , \\
    \sigma_{\text{FoG}} & \in [0, 20] \, , \\
    N & \in [-5000, 5000] \, .
\end{split}
\label{eq:prior}
\end{equation}
In eq.~\eqref{eq:likelihod} , $\mathbf{\Sigma}^{-1}$ is the inverse of the covariance matrix estimated with the EZmock realistic catalogs (see section~\ref{sec:mocks}). As a finite number of mocks, $N_{\text{m}} = 1000$, is used to estimate the covariance matrix, its inverse, the precision matrix, is biased \cite{2007A&A...464..399H}. This bias is corrected by re-scaling the covariance matrix with the inverse of the Hartlap factor,
\begin{equation}
    \mathbf{\Sigma'} = \frac{N_{\text{m}} - 1}{N_{\text{m}} - N_{\text{b}} - 2} \, \mathbf{\Sigma} \, ,
    \label{eq:wishart}
\end{equation}
where $N_{\text{b}}$ is the number of $k$-bins in the observed power spectrum. This correction is $\sim 5\%$ for the NGC and $\sim 6\%$ for the SGC.

The MonteCarlo Markov Chain (MCMC) algorithm employed in this analysis is the Hamiltonian MonteCarlo~\citep[HMC;][]{1987PhLB..195..216D,2011hmcm.book..113N,2017arXiv170102434B}.
HMC is a sampling technique that combines principles from Hamiltonian mechanics and MCMC methods to efficiently explore high-dimensional parameter spaces. Unlike traditional methods, HMC employs a dynamic integration of the target probability distribution. By introducing auxiliary momentum variables, HMC maps the trajectory of the walkers exploring the posterior distribution into an Hamiltonian system, whose potential energy is defined by the product of the likelihood and the prior distribution. Then, HMC exploits the gradients of the distribution to follow the Hamiltonian trajectories and efficiently sample the posterior even in a high-dimensional space.
In this work we use the No-U-Turn Sampler \citep[NUTS;][]{2011arXiv1111.4246H,ge2018t} implementation of HMC, which is able to automatically adapt critical parameters like the step size and the trajectory length.
By setting the acceptance rate to $0.9$, the chains quickly converge, within a few thousands steps, to $R - 1 \lesssim 10^{-3}$, where $R$ is the Gelman-Rubin statistics \citep{1992StaSc...7..457G}.

\section{Constraints and Discussion} \label{sec:results}

\begin{figure}
    \centering
    \includegraphics[width=.5\textwidth]{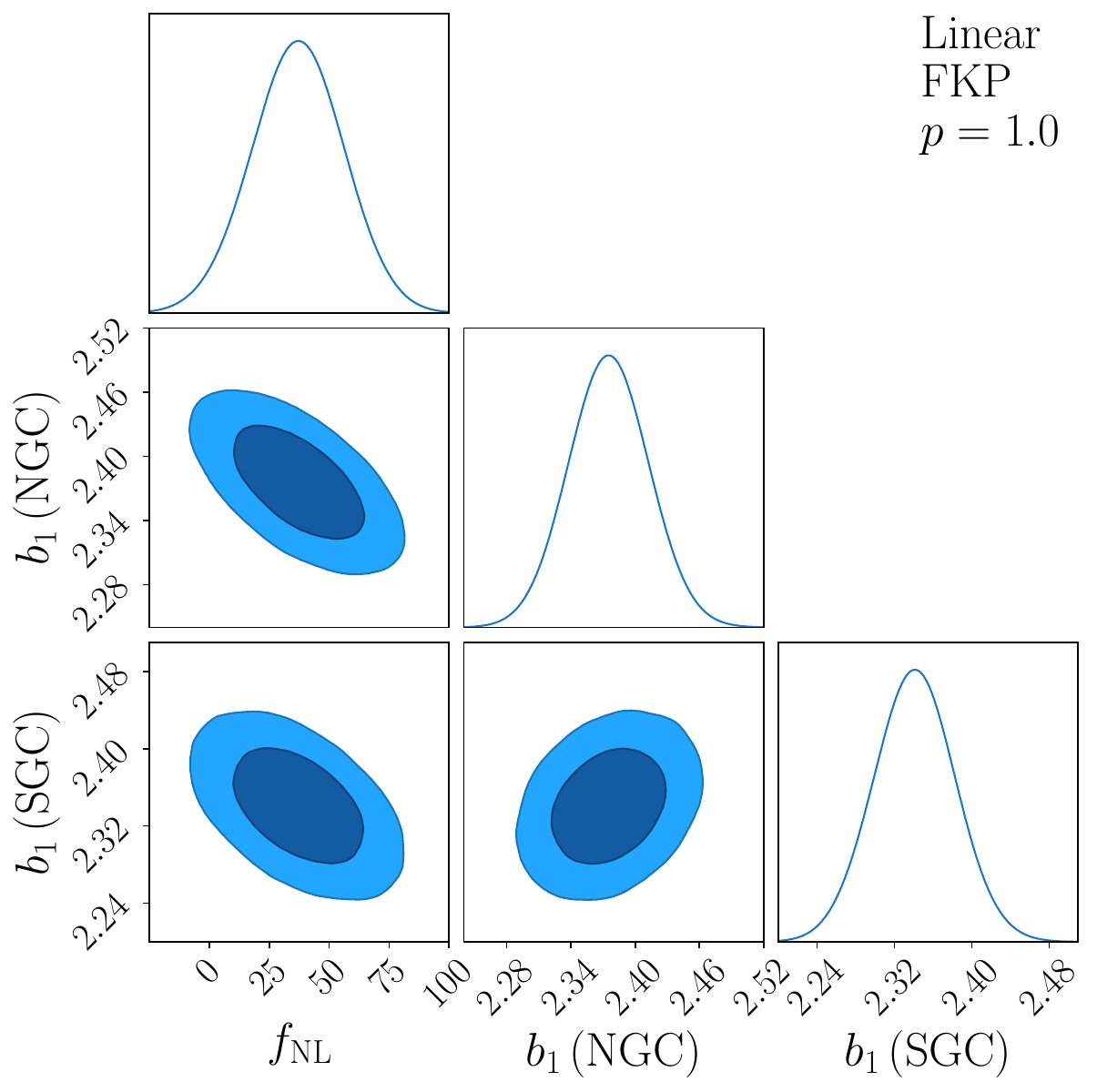}\includegraphics[width=.5\textwidth]{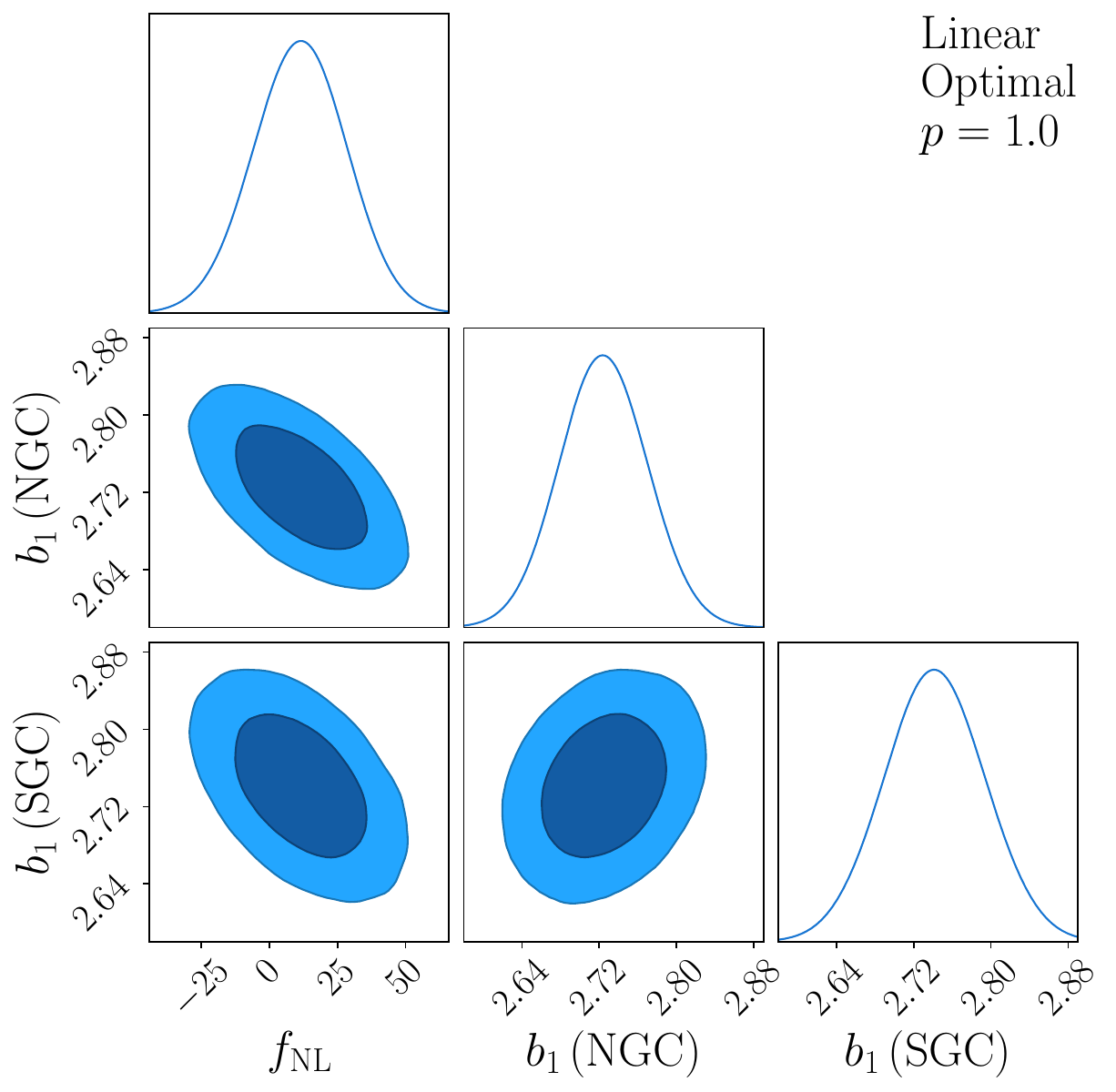}
    \includegraphics[width=.5\textwidth]{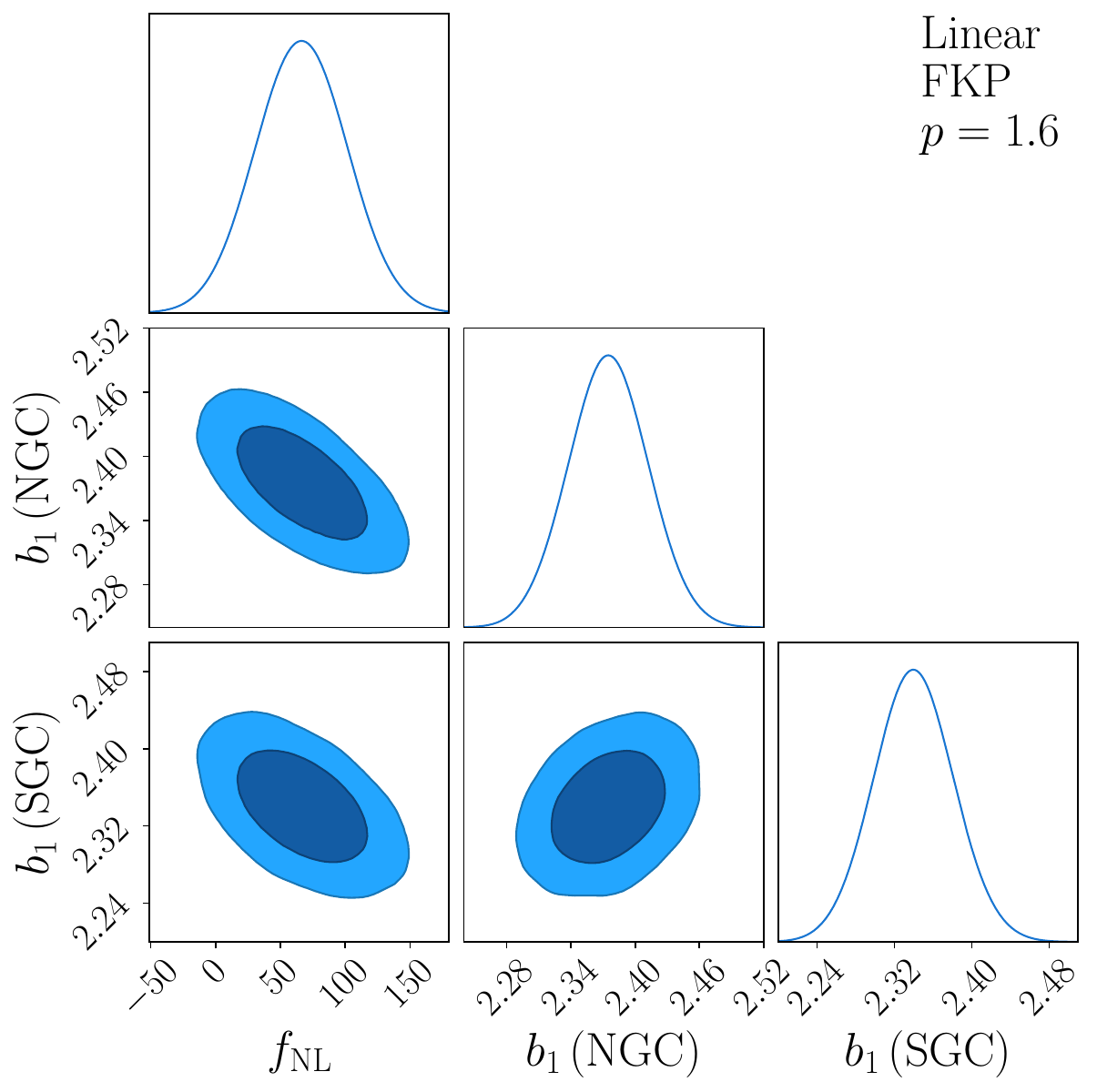}\includegraphics[width=.5\textwidth]{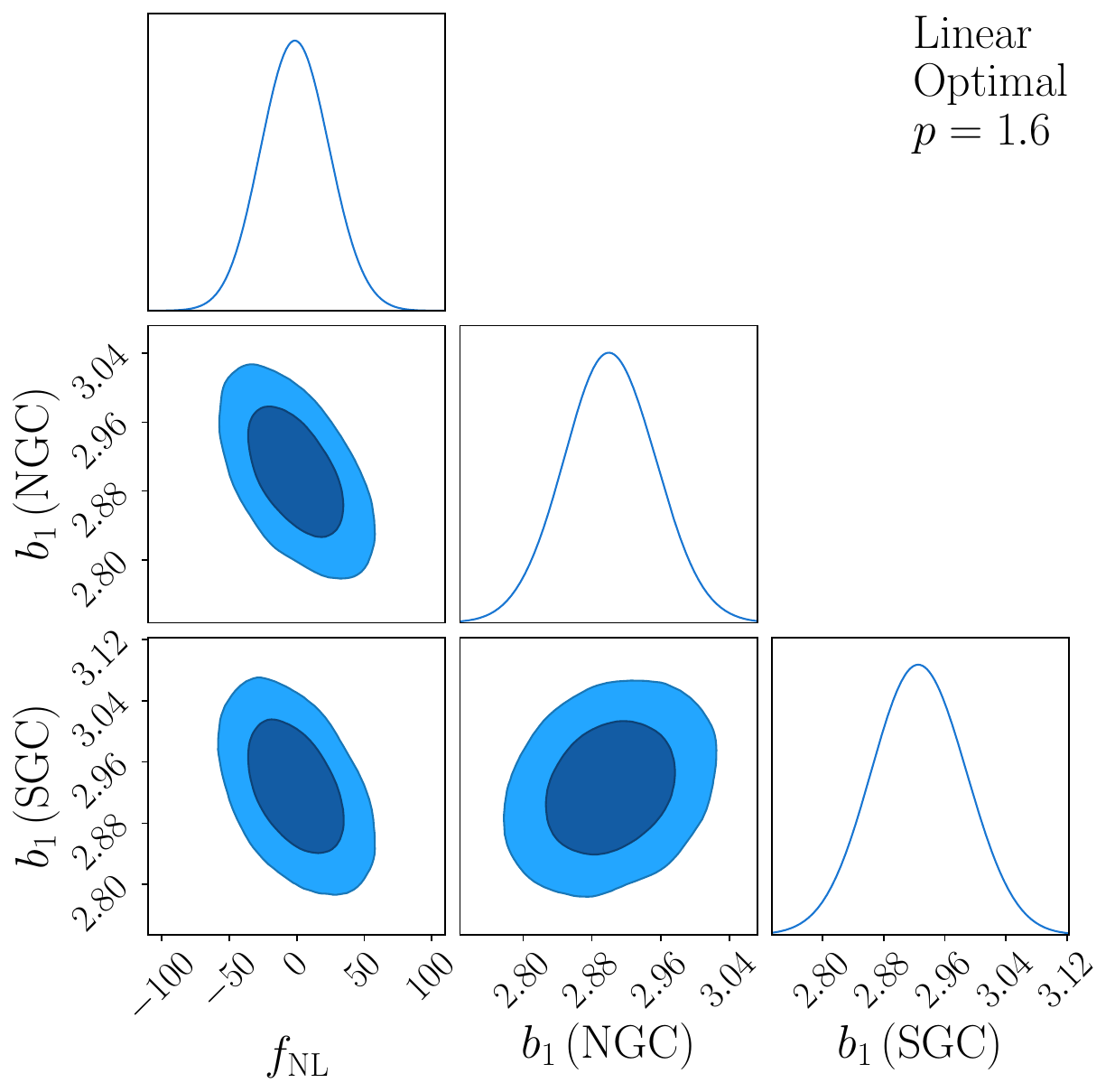}
    \caption{Two-dimensional posterior distributions for $f_{\text{NL}}$ and the quasar bias, $b_1$, of NGC and SGC from the joint analysis of the linear catalog. The plots on the top corresponds to $p=1.0$, and the bottom plots to $p=1.6$. On the left are the results for the FKP weighting scheme, and on the right for the optimal weights.}
    \label{fig:corner-linear}
\end{figure}
\begin{figure}
    \centering
    \includegraphics[width=.5\textwidth]{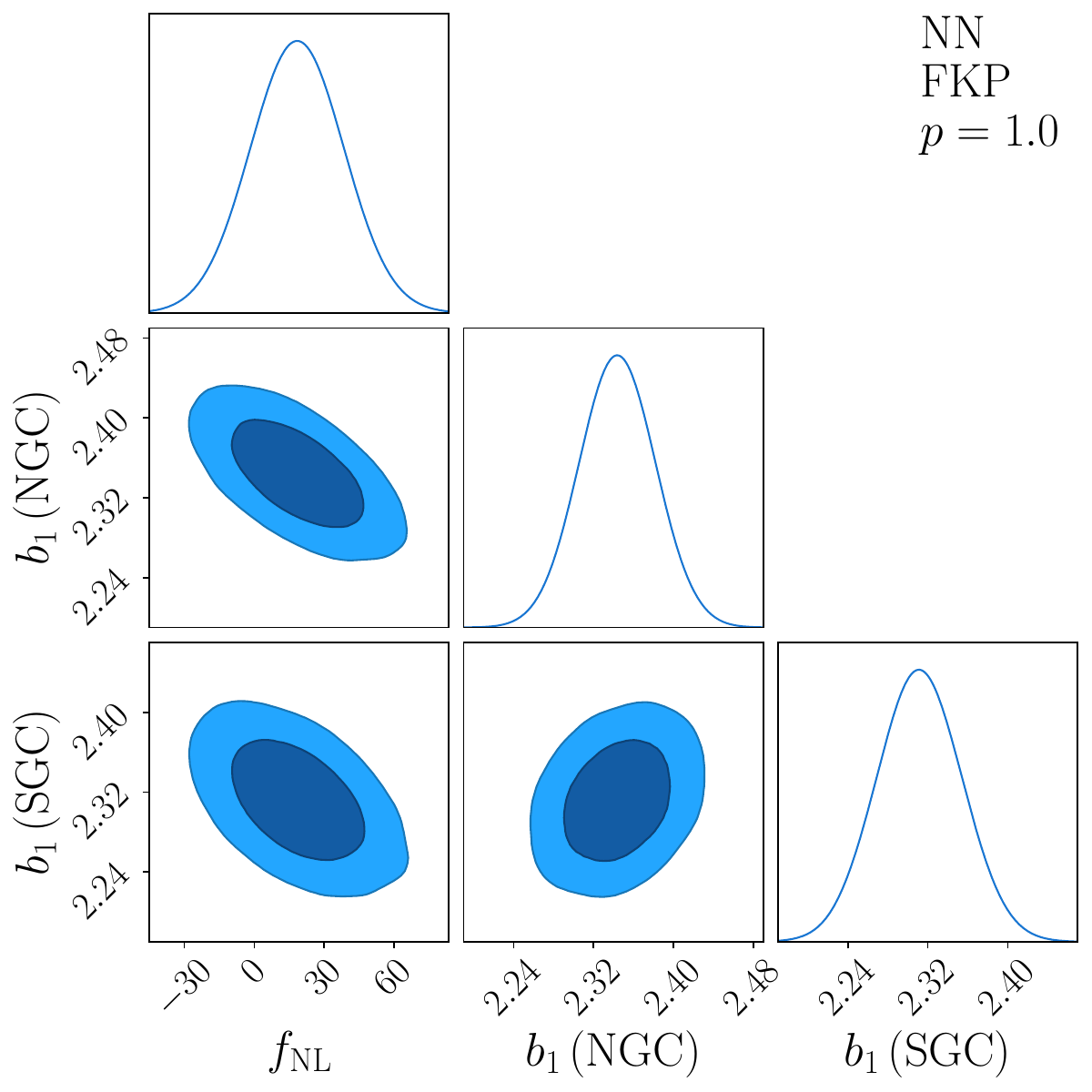}\includegraphics[width=.5\textwidth]{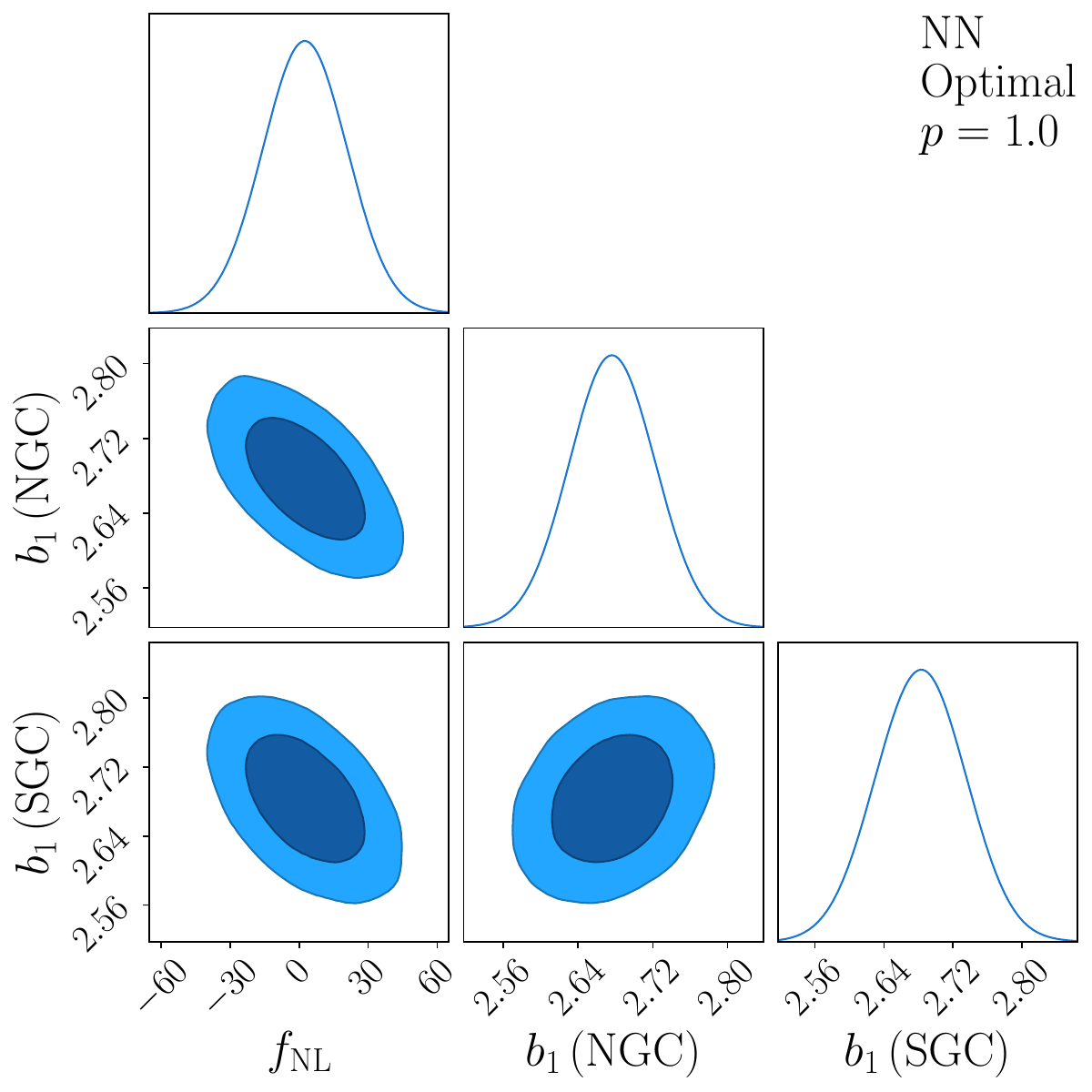}
    \includegraphics[width=.5\textwidth]{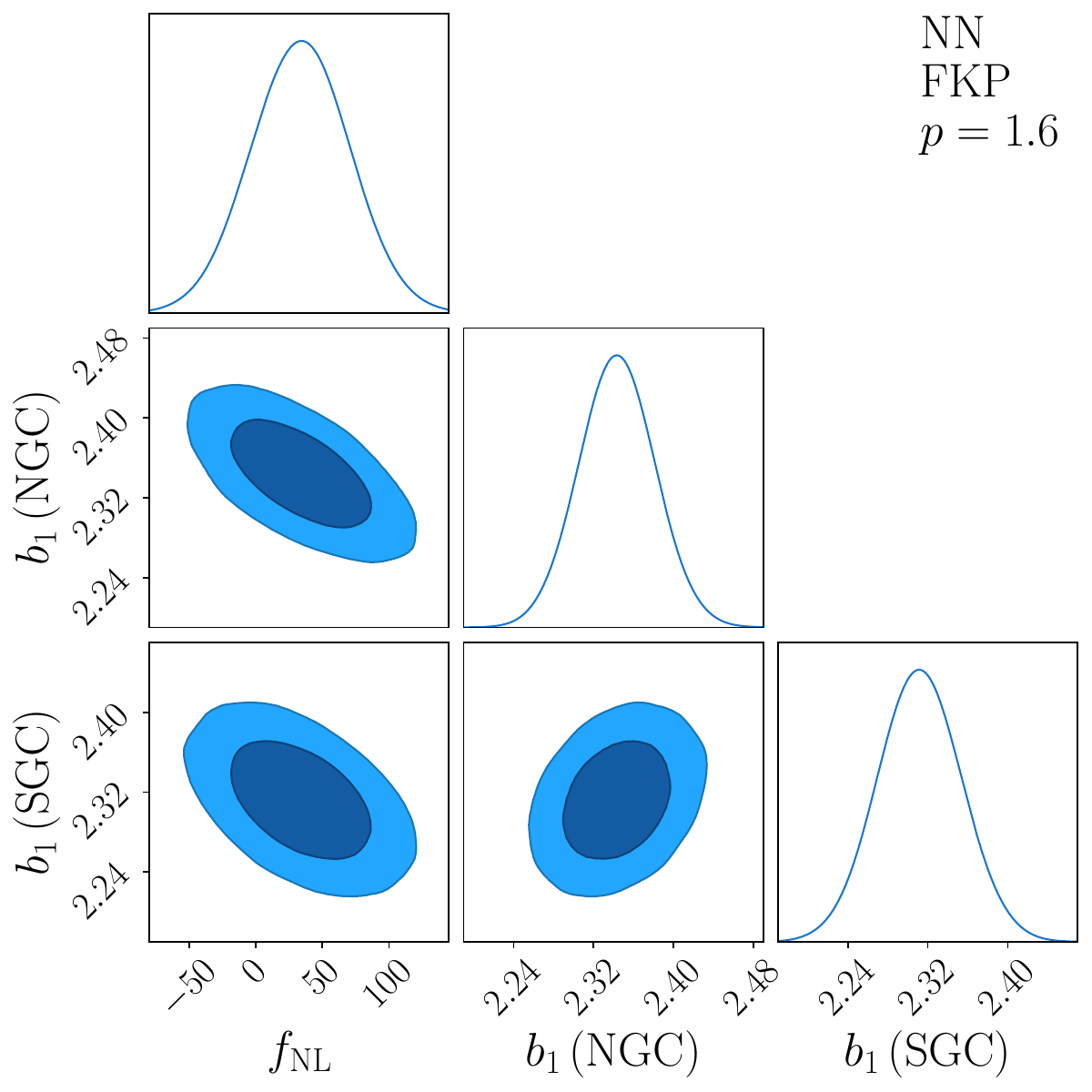}\includegraphics[width=.5\textwidth]{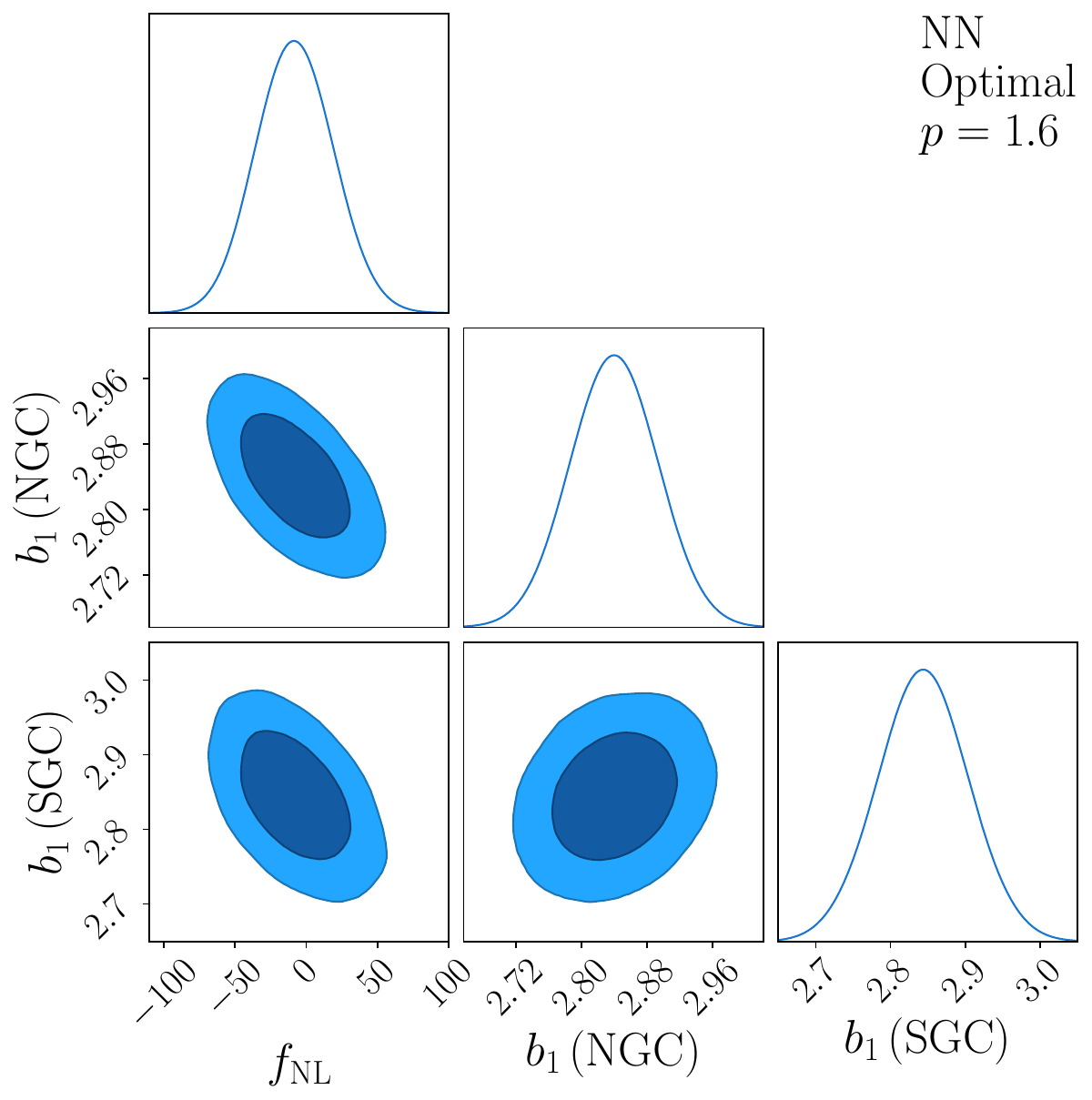}
    \caption{Same as figure~\ref{fig:corner-linear}, but for the NN catalog.}
    \label{fig:corner-NN}
\end{figure}
\begin{figure}
    \centering
    \includegraphics[width=.5\textwidth]{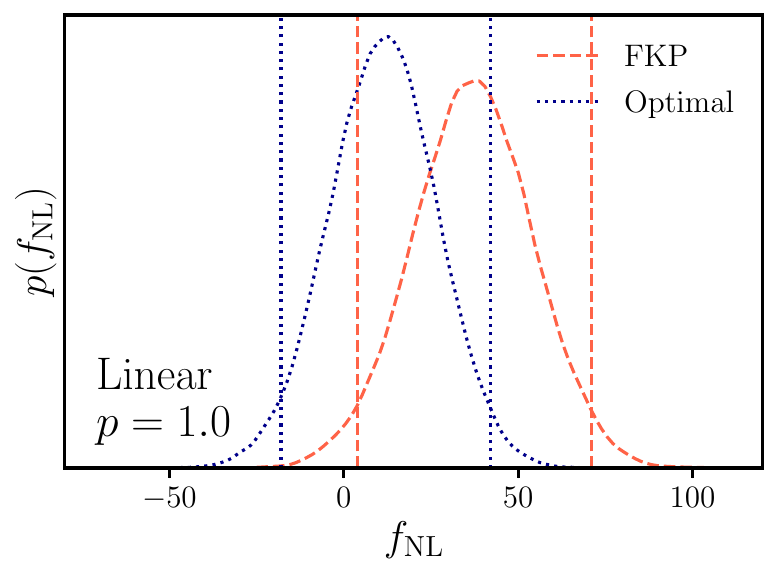}\includegraphics[width=.5\textwidth]{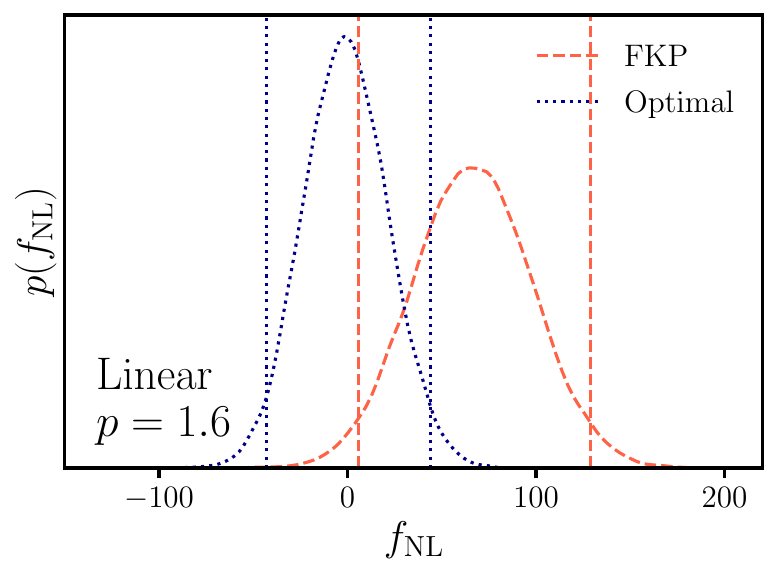}
    \caption{One dimensional posterior distribution for $f_{\text{NL}}$ for the joint analysis of the linear catalog. The dashed red curve is the posterior distribution obtained with the FKP weight analysis, and the dotted blue curve with the optimal weights analysis. The vertical lines mark the corresponding $95\%$ constraints. On the left, the results for $p=1.0$, and on the right for $p=1.6$.}
    \label{fig:fnl-posterior}
\end{figure}
\begin{table}[]
    \centering
    \begin{tabular}{ccccccc} \toprule
         & $p$ & & C.L. & Linear & NN & eBOSS DR14Q\\ \midrule
         \multirow{8}{2em}{joint} & \multirow{4}{1em}{$1.0$} & \multirow{2}{3em}{FKP} & $68\%$ & $19 < f_{\text{NL}} < 53 $ & $0 < f_{\text{NL}} < 36$ & \\ \cmidrule(l){4-7}
         & & & $95\%$ & $4 < f_{\text{NL}} < 71$ & $-16 < f_{\text{NL}} < 54$ & $-39 < f_{\text{NL}} < 41$ \\ \cmidrule(l){3-7}
         & & \multirow{2}{3em}{Optimal} & $68\%$ & $-4 < f_{\text{NL}} < 27$ & $-14 < f_{\text{NL}} < 19$ &  \\ \cmidrule(l){4-7}
         & & & $95\%$ & $-18 < f_{\text{NL}} <42$ & $-30 < f_{\text{NL}} < 34$ & $-51 < f_{\text{NL}} < 21$  \\ \cmidrule(l){2-7}
         & \multirow{4}{1em}{$1.6$} & \multirow{2}{3em}{FKP} & $68\%$ & $34 < f_{\text{NL}} < 97$ & $0 < f_{\text{NL}} < 66$ & \\ \cmidrule(l){4-7}
         & &  & $95\%$ & $6 < f_{\text{NL}} < 129$ & $-32 < f_{\text{NL}} < 98$ & $-74 < f_{\text{NL}} < 81$ \\ \cmidrule(l){3-7}
         & & \multirow{2}{3em}{Optimal} & $68\%$ & $-23 < f_{\text{NL}} < 21$ & $-33 < f_{\text{NL}} < 15$ & \\ \cmidrule(l){4-7}
         & & & $95\%$ & $-43 < f_{\text{NL}} < 44$ & $-54 < f_{\text{NL}} < 40$ & $-81 < f_{\text{NL}} < 26$  \\ \midrule \midrule

         \multirow{8}{2em}{NGC} & \multirow{4}{1em}{$1.0$} & \multirow{2}{3em}{FKP} & $68\%$ & $26 < f_{\text{NL}} < 67$ & $1 < f_{\text{NL}} < 44$ & \\ \cmidrule(l){4-7}
         & & & $95\%$ & $6 < f_{\text{NL}} < 87$ & $-23 < f_{\text{NL}} < 63$ & $-34 < f_{\text{NL}} < 61$ \\ \cmidrule(l){3-7}
         & & \multirow{2}{3em}{Optimal} & $68\%$ & $-6 < f_{\text{NL}} < 34$ & $-20 < f_{\text{NL}} < 21$ &  \\ \cmidrule(l){4-7}
         & & & $95\%$ & $-27 < f_{\text{NL}} < 62$ & $-42 < f_{\text{NL}} < 41$ & $-56 < f_{\text{NL}} < 38$  \\ \cmidrule(l){2-7}
         & \multirow{4}{1em}{$1.6$} & \multirow{2}{3em}{FKP} & $68\%$ & $49 < f_{\text{NL}} < 125$ & $-2 < f_{\text{NL}} < 78$ & \\ \cmidrule(l){4-7}
         & &  & $95\%$ & $10 < f_{\text{NL}} < 159$ & $-38 < f_{\text{NL}} < 121$ & $-67 < f_{\text{NL}} < 112$ \\ \cmidrule(l){3-7}
         & & \multirow{2}{3em}{Optimal} & $68\%$ & $-39 < f_{\text{NL}} < 28$ & $-53 < f_{\text{NL}} < 17$ & \\ \cmidrule(l){4-7}
         & & & $95\%$ & $-80 < f_{\text{NL}} < 54$ & $-95 < f_{\text{NL}} < 47$ & $-87 < f_{\text{NL}} < 42$  \\ \midrule \midrule

         \multirow{8}{2em}{SGC} & \multirow{4}{1em}{$1.0$} & \multirow{2}{3em}{FKP} & $68\%$ & $-15 < f_{\text{NL}} < 41$ & $-17 < f_{\text{NL}} < 45$ & \\ \cmidrule(l){4-7}
         & & & $95\%$ & $-36 < f_{\text{NL}} < 72$ & $-43 < f_{\text{NL}} < 78$ & $-64 < f_{\text{NL}} < 31$ \\ \cmidrule(l){3-7}
         & & \multirow{2}{3em}{Optimal} & $68\%$ & $-17 < f_{\text{NL}} < 29$ & $-21 < f_{\text{NL}} < 32$ &  \\ \cmidrule(l){4-7}
         & & & $95\%$ & $-35 < f_{\text{NL}} < 55$ & $-40 < f_{\text{NL}} < 63$ & $-61 < f_{\text{NL}} < 26$ \\ \cmidrule(l){2-7}
         & \multirow{4}{1em}{$1.6$} & \multirow{2}{3em}{FKP} & $68\%$ & $-22 < f_{\text{NL}} < 78$ & $-34 < f_{\text{NL}} < 81$ & \\ \cmidrule(l){4-7}
         & &  & $95\%$ & $-61 < f_{\text{NL}} < 135$ & $-82 < f_{\text{NL}} < 146$ & $-122 < f_{\text{NL}} < 63$ \\ \cmidrule(l){3-7}
         & & \multirow{2}{3em}{Optimal} & $68\%$ & $-28 < f_{\text{NL}} < 36$ & $-37 < f_{\text{NL}} < 40$ & \\ \cmidrule(l){4-7}
         & & & $95\%$ & $-51 < f_{\text{NL}} < 74$ & $-66 < f_{\text{NL}} < 84$ & $-92 < f_{\text{NL}} < 42$ \\          
         \bottomrule
    \end{tabular}
    \caption{Summary of the $f_{\text{NL}}$ $68\%$ and $95\%$ constraints of this work. The results for NGC, SGC and the joint analysis are presented, and compared with the eBOSS DR14Q constraints \citep{Castorina2019}.}
    \label{tab:fnl}
\end{table}
In this section we present and discuss the constraints on $f_{\text{NL}}$ we obtained with the analyses of DR16Q. Figure~\ref{fig:Pk-obs} shows the measured data points and error bars of the monopole of the power spectrum with the best fit model of the joint analyses. The plots are presented for the two sky region, NGC (left column) and SGC (right column); and the the three weighting schemes, the standard FKP with a model assuming $p=1.6$ (top row), the optimal weights for $p=1.0$ (middle row) and $p=1.6$ (bottom row). 

Figures~\ref{fig:corner-linear} and \ref{fig:corner-NN} show the two dimensional posterior of the joint analyses, respectively of the linear and NN catalog. Both figures are organized as follows: the left column corresponds to the analysis of the power spectrum monopoles measured with the standard FKP weights and the right column to the analysis of the optimally weighted power spectrum; the top row presents the results for $p=1.0$, and the bottom row to $p=1.6$. In the plots we only show three of the seven fit parameters: $f_{\text{NL}}$, and the linear bias $b_1$ of the two sky caps. In all the analyses of both the catalogs there is almost no correlation between the bias of the two sky region. This is expected as they corresponds to independent fields of view. Moreover, the linear bias of the two Galactic caps are always consistent with each other, and the bias estimated with the optimal weights are larger than the bias estimated using the standard FKP weights. The reason behind this behavior is the higher effective redshift of the sample when using the optimal weighting scheme, as discussed in section~\ref{sec:power_spec}. Another effect visible in figures~\ref{fig:corner-linear} and \ref{fig:corner-NN} is how the correlation between the linear bias and $f_{\text{NL}}$ changes between the FKP weights and the optimal weights. Even though this effect is present for both $p$ it is more evident in the case with $p=1.6$. The optimal weights reduce the correlation between $f_{\text{NL}}$ and the other parameters, here we are only showing the linear bias.

The comparison of the one-dimensional $f_{\text{NL}}$ posterior distributions obtained analyzing the linear catalog are shown in figure~\ref{fig:fnl-posterior}. The left panel corresponds to the case with $p=1.0$, and the right panel to $p=1.6$. In both panels the red dashed line is the $f_{\text{NL}}$ posterior of the FKP analysis, and the dotted blue line represents the posterior of the optimal weight analysis. The corresponding vertical lines mark the $95\%$ constraints. For both values of $p$ the $95\%$ constraints estimated with the standard FKP weights do not contain $f_{\text{NL}} = 0$. Given that CMB, which measures $\fnl = 0.8 \pm 5$, and LSS probe the primordial power spectrum over the same range of scales, this suggests the presence of residual contamination in the FKP catalog produced with the linear systematic weights.
The optimal weights shift the posterior to values more consistent with $f_{\text{NL}} = 0$. Serendipitously, this suggests that higher redshifts QSOs in the samples might be less affected by systematic effects. 
Nevertheless, the most important difference between the standard FKP and optimal weights is that the optimal weights give tighter constraints.  For $p=1.0$ the optimal weights improve the $95\%$ constraint of about $10\%$, for $p=1.6$ this improvement is a little less than $30\%$. 
The comparison between the weighted and the un-weighted constraints is difficult, due to the large systematic effects still present in the FKP catalogs, especially the NGC ones. At 95\% c.l. the joint analysis for $p=1.0$ does not contain $f_{\rm NL} = 0 $ for the linear catalogs, and it does only at the 68\% c.l. for the NN ones, see table~\ref{tab:fnl}. 
Nevertheless, the larger improvements with $p=1.6$ than with $p=1.0$ points in the direction of $b_{\phi} \sim b_1 - 1.6$. 

The $68\%$ and $95\%$ constraints on $f_{\text{NL}}$ are written down in table~\ref{tab:fnl}, where we also compare to the results 
of ref.~\cite{Castorina2019} on the eBOSS DR14Q data. The constraints from the NN catalog on $f_{\text{NL}}$ for a single sky patch are tighter than the ones from linear catalog. This was expected, as the first bin of the power spectrum monopole computed with the NN catalog is systematically lower than the monopole of the linear catalog. However, in the case of the joint analysis we always get tighter constraints with the linear catalog rather than with the NN catalog. The improvement in the constraints with the optimal weights with respect to the standard FKP ones is the same for the two catalogs. On the other hand, the improvements with respect to the eBOSS DR14Q analysis is between $10\%$ and $20\%$ for both the catalogs, and is smaller than the one expected by the doubling of the survey volume. This can be attributed to large scale systematic effects that are still present in the sample. Actually, when analyzing the single fields, the constraints show improvements with respect to eBOSS DR14Q that are smaller than the joint analysis, and in the case of SGC they are worse than the older analysis. Nevertheless, the single field analyses give posterior distributions consistent with each other and their combination in the joint analysis produces the tightest constraints. The best constraints of this work are from the joint analysis of the linear catalog with the optimal weights. The $95\%$ constraints for $p=1.0$ are $-18 < f_{\text{NL}} < 42$ corresponding to $\sigma_{f_{\text{NL}}} \sim 15$ and for $p=1.6$ they are $-43 < f_{\text{NL}} < 44$ with $\sigma_{f_{\text{NL}}} \sim 22$.

We have also repeated the parameters estimation assuming a value of $p=3.0$ in the optimal weights. In this case, we expect worse constraint on $f_{\rm NL}$ from both the FKP and the optimal analysis compared to the cases discussed above. However, as discussed in section~\ref{sec:introduction}, we can use these constraints to show how the optimal analysis can provide a data-driven estimate of the value of $p$. If the value of $p$ used in the optimal weights is not close to the true one, then the optimal analysis will not improve over the standard case, or will improve less than an analysis with a value of $w_0\sim b_\phi$ closer to the actual response. Note that this approach assumes that the signal one is looking for is non-zero.
We test this idea for $p=3.0$. In this case, the NGC analysis always shows evidence for non-zero $f_{\rm NL}$ at the $95\%$ c.l., thus we focus always on SGC. For the linear catalog, the constraints are $-140< f_{\text{NL}} < 81$ for the FKP weights and $-129 < f_{\text{NL}} < 76$ for the optimal analysis of the NN catalog. This improvement by $8\%$ should be compared to the $20\%$ and $56\%$ reduction of the error bar for $p=1.0$ and $p=1.6$ respectively. This implies that larger values of $p\gtrsim3$ are disfavored for this sample.

\section{Conclusions} \label{sec:conclusions}
In this work we presented the most stringent constraint on the amplitude of local Primordial Non-Gaussianities with Large Scale Structure data, in particular with the eBOSS DR16Q data set. Assuming the QSOs response to $f_{\rm NL}$ is proportional to $b_1-p$, where $b_1$ is the linear bias, our strongest bounds read
\begin{equation}
\begin{split}
    -4 < f_{\text{NL}} < 27 \, ,  &\quad \text{for} \; p=1.0 \, , \\
    -23 < f_{\text{NL}} < 21 \, , & \quad \text{for} \; p=1.6 \, ,
\end{split}
\end{equation}
at 68\% c.l.. 

Our goal was to show that the optimal signal weighting reduces the error bars on $f_{\rm NL}$ compared to a standard analysis, and we robustly find improvement between 10-30\% depending on the analysis setup.  While our optimal constraints are always consistent with no local PNG, the comparison with previous eBOSS data releases does not allow us to exclude the presence of residual systematic effects in the data. Nevertheless, the DR16Q catalog and the analysis presented here represent an important step forward in the direction of robust and optimal analysis of PNG with LSS data. 
We have also shown how optimal weights could provide a data-driven prior on the largely unknown value of $p$, and we were able to exclude $p \gtrsim 3$.

This work can be extended in several directions. First,  we have not attempted an optimal noise weighting of the power spectrum data. This could be done using optimal quadratic estimators \cite{1998ApJ...499..555T}, for which new algorithms have been recently presented \cite{Philcox:2020vbm}. A fully optimal analysis will allow to get closer to the full Fisher information contained in the power spectrum. Secondly, it is well known that the Bispectrum is the most sensitive probe to local PNG. A careful study of the optimal weights for higher-point statistics is still missing and could revolutionize the way we constrain $f_{\rm NL}$. We intend to return to these interesting problems in future work.

\acknowledgments
We thank Ashley Ross, Hector Gil-Marin, Arnaud de Mattia, Jiamin Hou and Eva-Maria Mueller for discussion on the use of eBOSS data. 

\bibliographystyle{JHEP}
\bibliography{main}

\appendix

\section{Fitting $b_{\phi} \, f_{\text{NL}}$}
\label{sec:App}
\begin{figure}
    \centering
    \includegraphics[width=.5\textwidth]{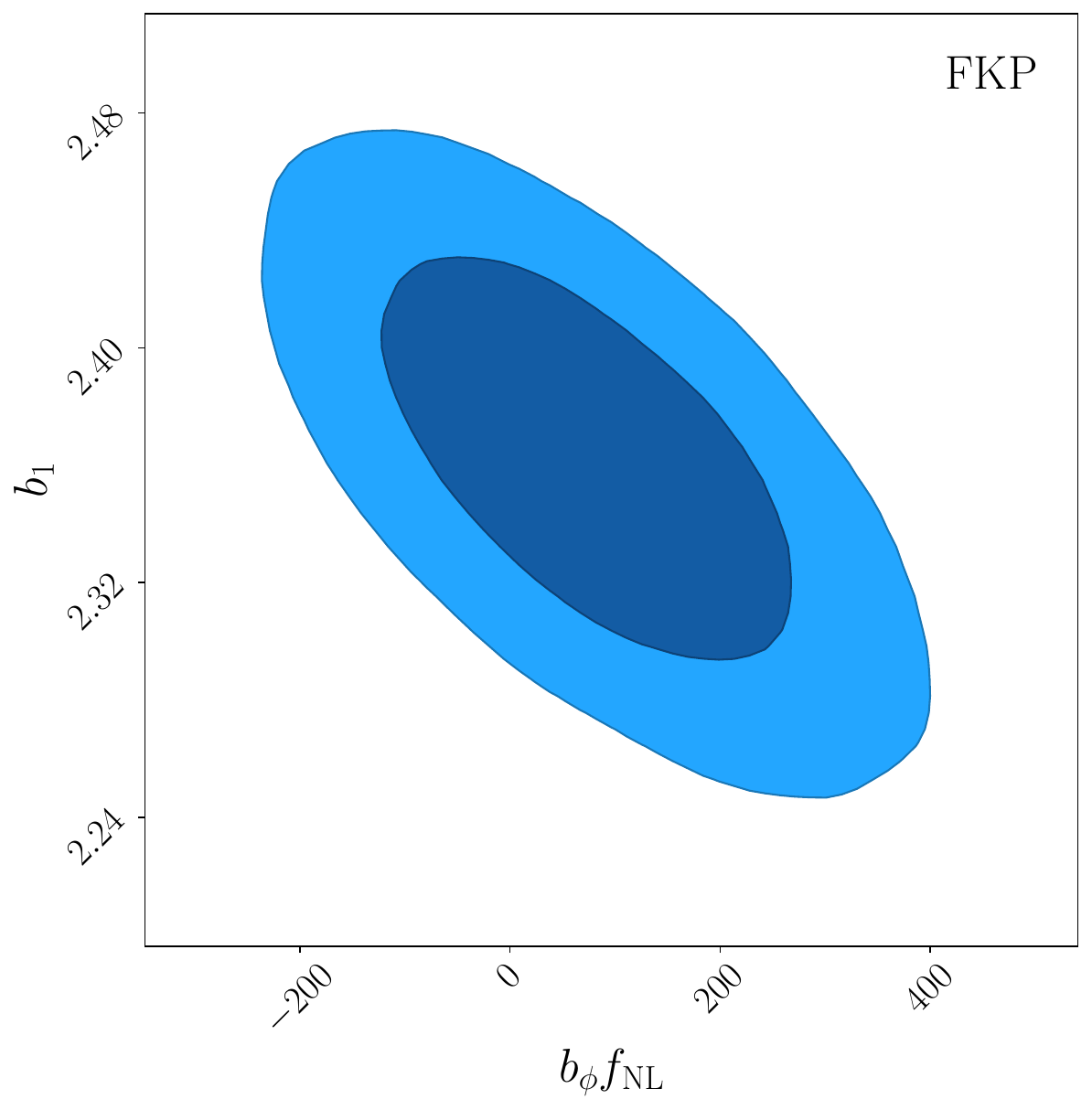}\includegraphics[width=.5\textwidth]{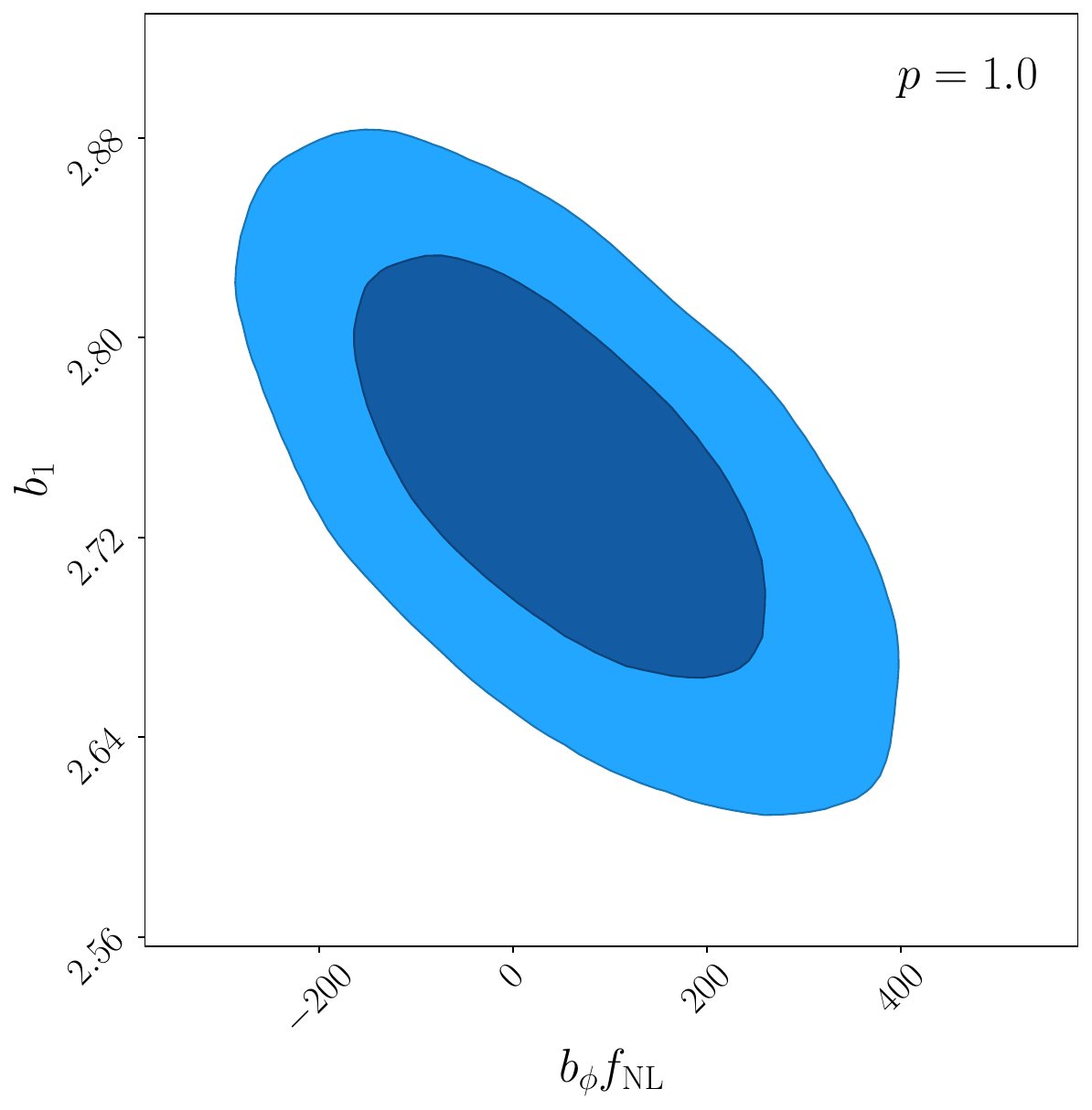}
 \includegraphics[width=.5\textwidth]{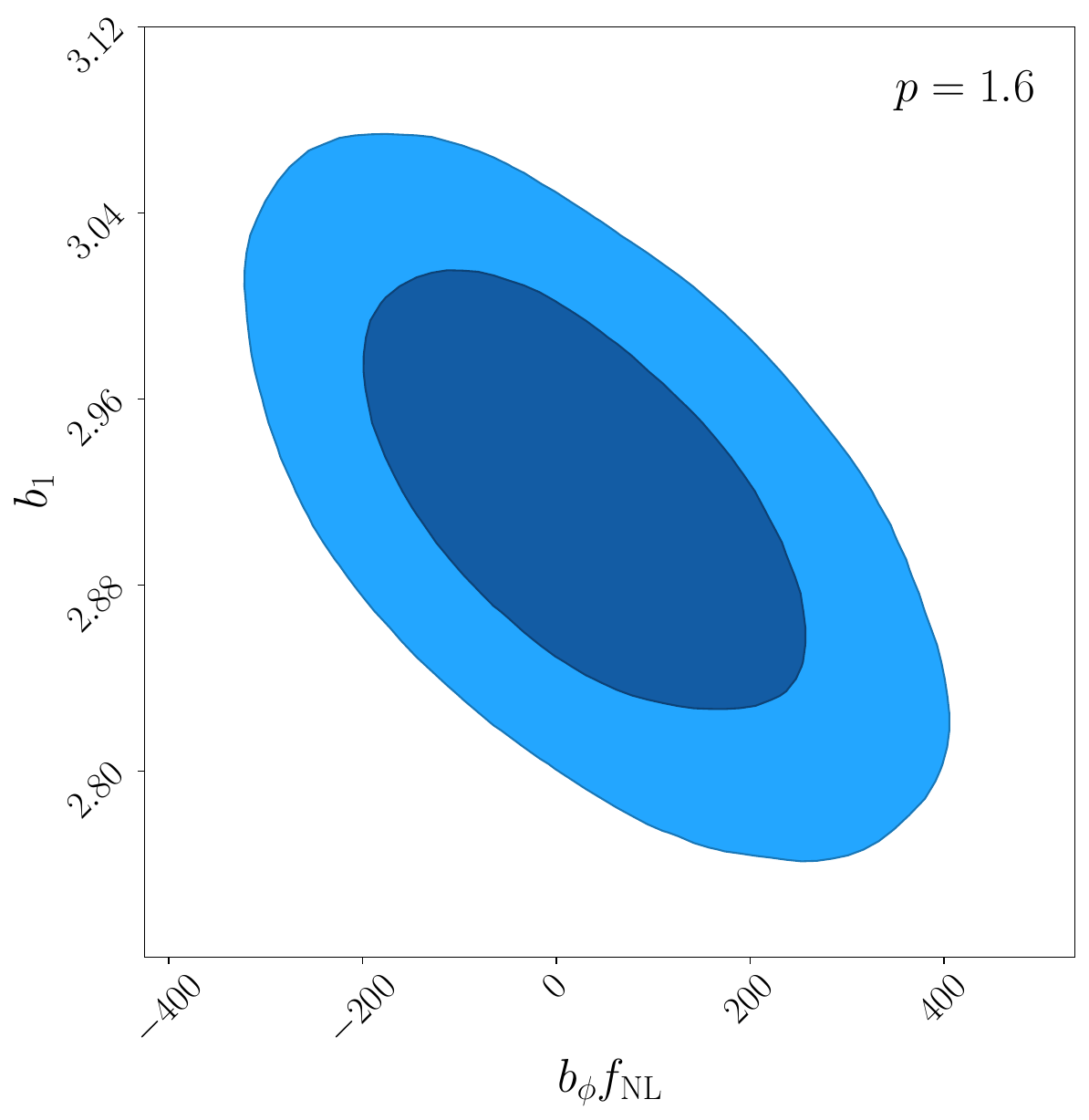}
    \caption{Two dimensional posterior distributions for $b_{\phi} \, f_{\text{NL}}$ and and the quasar linear bias, $b_1$, from the SGC linear catalog analysis. The posterior distribution of the analysis with the FKP weighting scheme, the optimal weights with $p=1.0$, and $p=1.6$ are shown respectively on the top left, top right and bottom panel.}
    \label{fig:bphi-fnl}
\end{figure}

Table~\ref{tab:bphi-fnl} summarizes the constraint on $b_\phi f_{\rm NL}$ for the different measurements of the power spectrum. As in the main text, all bounds are compatible with zero PNG with the exception of the FKP analysis of NGC. 
Figure~\ref{fig:bphi-fnl} shows the 2D posterior of the parameters in SGC.
\begin{table}[]
    \centering
    \begin{tabular}{ccccc} \toprule
         & & C.L. & Linear & NN \\ \midrule
         \multirow{6}{2em}{NGC} & \multirow{2}{4em}{FKP} & $68\%$ & $ 131 < b_{\phi} \, f_{\text{NL}} < 310$ & $2 < b_{\phi} \, f_{\text{NL}} < 192$ \\ \cmidrule(l){3-5}
         & & $95\%$ & $35 < b_{\phi} \, f_{\text{NL}} < 394$ & $-102 < b_{\phi} \, f_{\text{NL}} < 280$ \\ \cmidrule(l){2-5}
         & \multirow{2}{4em}{Optimal $p=1.0$} & $68\%$ & $-27 < b_{\phi} \, f_{\text{NL}} < 200$ & $-115 < b_{\phi} \, f_{\text{NL}} < 123$ \\ \cmidrule(l){3-5}
         & & $95\%$ & $-174 < b_{\phi} \, f_{\text{NL}} < 288$ & $-255 < b_{\phi} \, f_{\text{NL}} < 227$ \\ \cmidrule(l){2-5}
         & \multirow{2}{4em}{Optimal $p=1.6$} & $68\%$ & $-176 < b_{\phi} \, f_{\text{NL}} < 121$ & $-222 < b_{\phi} \, f_{\text{NL}} < 76$ \\ \cmidrule(l){3-5}
         & & $95\%$ & $-371 < b_{\phi} \, f_{\text{NL}} < 234$ & $-409 < b_{\phi} \, f_{\text{NL}} < 192$ \\ \midrule \midrule

        \multirow{6}{2em}{SGC} & \multirow{2}{4em}{FKP} & $68\%$ & $-48 < b_{\phi} \, f_{\text{NL}} < 198$ & $-83 < b_{\phi} \, f_{\text{NL}} < 188$ \\ \cmidrule(l){3-5}
         & & $95\%$ & $-162 < b_{\phi} \, f_{\text{NL}} < 322$ & $-207 < b_{\phi} \, f_{\text{NL}} < 323$ \\ \cmidrule(l){2-5}
         & \multirow{2}{4em}{Optimal $p=1.0$} & $68\%$ & $-93 < b_{\phi} \, f_{\text{NL}} < 175$ & $-129 < b_{\phi} \, f_{\text{NL}} < 174$ \\ \cmidrule(l){3-5}
         & & $95\%$ & $-202 < b_{\phi} \, f_{\text{NL}} < 317$ & $-230 < b_{\phi} \, f_{\text{NL}} < 349$ \\ \cmidrule(l){2-5}
         & \multirow{2}{4em}{Optimal $p=1.6$} & $68\%$ & $-121 < b_{\phi} \, f_{\text{NL}} < 165$ & $-144 < b_{\phi} \, f_{\text{NL}} < 182$ \\ \cmidrule(l){3-5}
         & & $95\%$ & $-237 < b_{\phi} \, f_{\text{NL}} < 311$ & $-265 < b_{\phi} \, f_{\text{NL}} < 354$ \\ \bottomrule
    \end{tabular}
    \caption{Summary on the $68\%$ and $95\%$ constraints on $b_{\phi} \, f_{\text{NL}}$ for the NGC and SGC.}
    \label{tab:bphi-fnl}
\end{table}

\end{document}